\documentclass[journal,12pt,onecolumn,draftclsnofoot]{IEEEtran}

\usepackage[utf8]{inputenc} 
\usepackage{cite}
\usepackage{graphicx}
\usepackage{epstopdf}
\usepackage[cmex10]{amsmath}
\usepackage{balance}
\usepackage{array}
\usepackage[caption=false,font=footnotesize]{subfig}
\usepackage{url}
\usepackage{amsfonts}
\usepackage{eurosym}
\usepackage{tabularx}
\usepackage{amsmath}
\usepackage{epstopdf}
\usepackage{changepage}
\usepackage{suffix}
\usepackage{mathtools}
\usepackage{multirow}

\IEEEoverridecommandlockouts

\begin{document}

\title{A Tractable Closed-Form Approximation of the Ergodic Rate in Poisson Cellular Networks}

\author{Alexis I. Aravanis (AA)$^{1*}$, Thanh Tu Lam (TTL)$^2$, Olga Muñoz$^1$ (OMM), Antonio Pascual-Iserte (API)$^1$, Marco Di Renzo (MDR)$^2$\\$^1$Department of Signal Theory and Communications, Universitat Politècnica de Catalunya (UPC), Barcelona, Spain\\$^2$Laboratoire des Signaux et Systèmes, CNRS, CentraleSupélec, Univ. Paris-Sud, Universite Paris-Saclay, 91192 Gif-sur-Yvette, France\\ alexios.aravanis@upc.edu, lamthanh.tu@l2s.centralesupelec.fr, olga.munoz@upc.edu, marco.direnzo@l2s.centralesupelec.fr
\thanks{The work presented in the present paper has been carried out within the framework of the project ETN-5Gwireless (this project has received funding from the European Union’s Horizon 2020 research and innovation programme under the Marie Skłodowska-Curie grant agreement No. 641985). Moreover, the work has been partially funded through the grant 2017 SGR 578 (funded by the Catalan Government—Secretaria d’Universitats i Recerca, Departament d’Empresa i Coneixement, Generalitat de Catalunya, AGAUR) and the project TEC2016-77148-C2-1-R (AEI/FEDER, UE): 5G\&B-RUNNER-UPC (funded by the Agencia Estatal de Investigacion, AEI, and Fondo Europeo de Desarrollo Regional, FEDER).\\

DOI: 10.1186/s13638-019-1499-9}}

\markboth{Accepted paper at EURASIP Journal on Wireless Communications and Networking (2019) 2019:187}
{}

\maketitle

\begin{abstract} 
The employment of stochastic geometry for the analysis and design of ultra dense networks (UDNs) has provided significant insights into network densification. In addition to the characterization of the network performance and behavior, these tools can also be exploited toward solving complex optimization problems that could maximize the capacity benefits arising in UDNs. However, this is preconditioned on the existence of tractable closed form expressions for the considered figures of merit. In this course, the present paper introduces an accurate approximation for the moment generating function (MGF) of the aggregate other-cell interference created by base stations whose positions follow a Poisson point process of given spatial density. Given the pivotal role of the MGF of the aggregate interference in stochastic geometry and the tractability of the derived MGF, the latter can be employed to substantially simplify ensuing stochastic geometry analyses. Subsequently, the present paper employs the introduced MGF to provide closed form expressions for the downlink ergodic capacity for the interference limited case, and validates the accuracy of these expressions by the use of extensive Monte Carlo simulations. The derived expressions depend on the density of users and base stations, setting out a densification road map for network operators and designers of significant value.
\end{abstract}




%



\section{Methods/Experimental}
\label{sec:method}
The methods used in the present paper are based on the mathematical tools of random spatial processes and stochastic geometry. The analytical framework developed for the performance evaluation of ultra dense networks is validated by extensive Monte Carlo simulations.

\section{Introduction}
\label{sec:introduction}
The advent of multimedia interactive services and the surge in the number of interconnected devices has imposed the investigation of new approaches able to enhance wireless capacity in 5G networks. In this course, three prime axes of network flexibility have been leveraged, namely the employment of wider spectrum, the enhancement of spectral efficiency and the employment of smaller cell sizes and, thus, of smaller transmit distances. In retrospect over the evolution of wireless networks, the efficient spatial reuse of the spectrum, through the reduction of the inter-site distances, has provided, out of these three axes of flexibility, the most substantial capacity gains by a large margin \cite{History}. Hence, the densification of networks arises as the most prominent candidate for achieving the envisaged capacity increase in the 5G era as well.

In the direction of densifying their networks, network operators employ system level simulations and network measurements. However, over the last decade, the seminal work of Baccelli et al. \cite{Baccelli} gave rise to stochastic geometry as a tractable tool for the large-scale analysis and design of wireless networks. Indicatively, the formulation of tractable mathematical expressions for the expectation of the aggregate interference in wireless networks \cite{Haenggi} $-$which was not analytically formulated hitherto$-$ paved the way for the theoretical analysis of the performance of wireless networks. Thenceforth, a multitude of research works exploited these tools to provide significant insights into network densification. These insights are essential for understanding the innate features of dense networks and can be employed by network operators as densification road maps and guidelines for the use of system level simulators and of auxiliary network planning tools.

The insights provided by such theoretical analyses brought about significant changes in the understanding of wireless networks. In particular, for single-slope path loss models and for networks comprising significantly more users than base stations (BSs), it has been demonstrated that the user signal quality is independent of the BS density \cite{Andrews1}. Moreover, the probability of coverage (which constitutes the complementary cumulative distribution function (CCDF) of the signal to interference plus noise ratio (SINR)) is independent of the BS density and of the number of tiers \cite{Andrews2}. Similar analysis for the uplink (UL) has demonstrated that the UL signal to interference ratio (SIR) is also invariant of the BS density \cite{TWC}. As a result, it has been demonstrated that the network capacity increases linearly with the density of BSs and with the number of tiers \cite{Haenggi}.

These conclusions, however, which indeed hold for sparse wireless networks (e.g. tier of macro cells (MCs)), do not hold for extremely high BS densities. The reason for that is that after a BS densification threshold, the inter-site distances become so small that the proximity of the neighboring BSs allows them to create  line of sight (LOS) interference to the intended user. As a result, after this densification threshold, the probability of coverage is diminished precipitately due to the presence of LOS interference \cite{Marco1,Lopez1}.

As opposed to this behavior of networks comprising much more users than BSs, the system performance is not bounded by the aforementioned threshold in the case of networks with more BSs than users; which is, in fact, the case of the envisaged UDNs \cite{UDNs}. The reason for that is that the excess BSs that do not serve any user can be switched off, thus, reducing the system energy consumption and interference. In this setup, the probability of coverage will still decrease when LOS interference first appears in the network, however, as the BS density increases past this point, the probability of coverage will increase again with the density of BSs. That is, since for a fixed number of users the excess BSs will remain idle, and therefore will not create any interference, regardless of the density of BSs \cite{Lopez2}. 

Based on this comprehensive analysis, it has become evident that two key factors need to be taken into account by network operators in order to tap the capacity potential of UDNs. Firstly, the detrimental effect of LOS interference and, secondly, the beneficial effect of idle (i.e. non transmitting) BSs. The incorporation of these two effects in the design of UDNs could engender extraordinary capacity gains. In particular, the investigation of BS coordination schemes to counteract LOS interference could, indeed, allow for a linear capacity increase with the density of BSs. Besides, the coordination of only LOS interferers, able to communicate and coordinate directly with the intended user, would not require intricate coordination schemes. Additionally, leveraging on the beneficial effect of the idle mode, users could be clustered dynamically under a single BS, not necessarily the one providing the best service to each user. Thus, BSs that were acting as sources of principal interference to the network could be switched off. Given the high density of BSs, the connection to a neighboring BS after the best serving BS has been switched off would entail a minimal path loss increase, that would be outweighed by the interference mitigation gain achieved by strategically switching off BSs.

The development of mathematical frameworks for complex optimization problems, like the ones mentioned above poses a great challenge. The reason for that is that the majority of the stochastic geometry approaches in the literature, including the theoretical analyses presented above, involve intractable integrations. Even though such integrals can be computed numerically, allowing the analysis of the network behavior, they cannot be employed for the investigation of complex optimization problems. In these cases it is imperative that the considered objective functions, that evaluate the system performance, involve tractable closed form expressions. In this course, it is essential to exploit the available stochastic geometry tools to develop tractable and accurate approximations in addition to the available exact but cumbersome expressions. 

In this direction, the authors have already provided approximate but accurate closed form expressions for the DL rate of coordinating BSs \cite{VTC}, providing a figure of merit for the investigation of coordination schemes among BSs. However,  the derivation of closed form expressions for the DL ergodic rate of non coordinating BSs had remained an open issue in the literature hitherto. Addressing this gap in the literature, the present paper provides approximate but extremely accurate closed form expressions for the DL ergodic rate for the interference limited case. The expressions account for networks comprising more users than BSs, as well as networks comprising more BSs than users. In the latter case, the DL rate is associated by a closed form expression to the density of BSs and of users allowing for the investigation of complex optimization problems like the one mentioned above. Last but not least, in the direction of obtaining the aforementioned expressions, an extremely accurate and simple approximation of the MGF of the aggregate other-cell interference in the DL is provided. In the existing literature, the complexity of the exact MGF imposed inherent limitations on the extension of the stochastic geometry analysis to complex optimization problems. However, the present paper lifts these inherent limitations by introducing a simple, albeit extremely accurate expression for the MGF of the other cell interference.

To elaborate, the main contributions of the present paper can be summarized as follows:
\begin{itemize}
\item Derivation of a tractable and extremely accurate approximation for the MGF of the aggregate other-cell interference. The aggregate interference is created by BSs whose spatial distribution follows a homogeneous Poisson point process (PPP).
\item Extension of the MGF to account for two different scenarios. A realistic scenario where the fast fading characterizing the channels of the interferers is not known to the UE, and a scenario of academic interest, where the fast fading of the interferes can be taken into account.
\item Derivation of a tractable and accurate approximation for the coverage probability in the interference limited and fully loaded case (i.e. for $\lambda_{UE}\gg \lambda$).
\item Derivation of a tractable and accurate approximation for the DL ergodic rate, in closed form, in the interference limited and fully loaded case.
\item Derivation of a tractable and accurate approximation for the coverage probability in the interference limited, but not fully loaded case (i.e. the case of the envisaged UDNs). As opposed to the fully loaded case, the coverage probability depends on the density of users $\lambda_{UE}$ and the density of BSs $\lambda$.
\item Derivation of a tractable and accurate approximation for the DL ergodic rate, in closed form, in the interference limited, but not fully loaded case. The closed form of the expressions and their dependence on the density of users $\lambda_{UE}$ and BSs $\lambda$, constitutes an invaluable figure of merit for network planning and network densification.
\item The accuracy of the proposed approximations is validated through extensive Monte Carlo simulations, and the practicality of the derived expressions for future applications is outlined. 
\end{itemize}

The remainder of the paper is organized as follows. Section \ref{sec:Architecture} presents the considered network architecture. Section \ref{sec:MGF} presents the approximation for the MGF of the aggregate other-cell interference in the DL. Section \ref{sec:Ergodic} introduces for the first time in the literature closed form expressions for the ergodic rate in the DL for the interference limited case. Moreover, extensive simulation results are presented corroborating the accuracy of the derived expressions. Finally, Section \ref{sec:Conclusions} concludes the paper and presents perspectives.

\section{The Wireless Cellular Network }
\label{sec:Architecture}

A wireless cellular system is considered, comprising a set of BSs, denoted by $BS_i$, whose positions $x_i \in \mathbb{R}^2$ follow a spatial distribution given by a homogeneous PPP $\Psi$ of density $\lambda$ $(BSs / m^2)$.  Moreover, the positions of the overlaid user equipment (UE) follow also a spatial distribution given by a homogeneous PPP $\Phi$ of density $\lambda_{UE}$ $(UE/m^2 )$. The reference UE, denoted by ${U\!E}_0$, is located at the origin and is served by its closest BS, denoted by $BS_0$. The UE adjoined at the origin can be singled out and the location of the other UE follows the reduced Palm distribution of $\Phi$, which is the same as the original distribution $\Phi$ (as stated by Slivnyaks theorem \cite{Stoyan}). Hence, adjoining ${U\!E}_0$ at the origin does not change the distribution of $\Phi$.

For the sake of simplicity in the notation, it is assumed that all UE and BSs are equipped with one antenna and all BSs transmit at the same power level. Intra cell users are assumed to be sharing orthogonal resources, as is typically the case in the literature \cite{TWC}, whereas all BSs use the same frequency band. If not explicitly mentioned otherwise (as will be done in the following sections), it is assumed that the network comprises significantly more users than BSs (i.e. $\lambda_{UE}\gg \lambda$). As a result, every BS is active and transmitting, acting as an interferer in the DL. The other-cell interference in the DL is mathematically defined as the interference coming from all BSs residing at a distance $||x_i||$ from the origin greater than the distance $||x_0||$ of $BS_0$ from the origin, where $||\cdot||$ denotes the $l^2$-norm. Note that the origin is where the reference UE ${U\!E}_0$ is located. Last but not least, a single-slope unbounded path-loss model is assumed in the analysis.

\begin{figure}[h!]
  \begin{center}
   \centerline{
    \includegraphics[height=9cm]{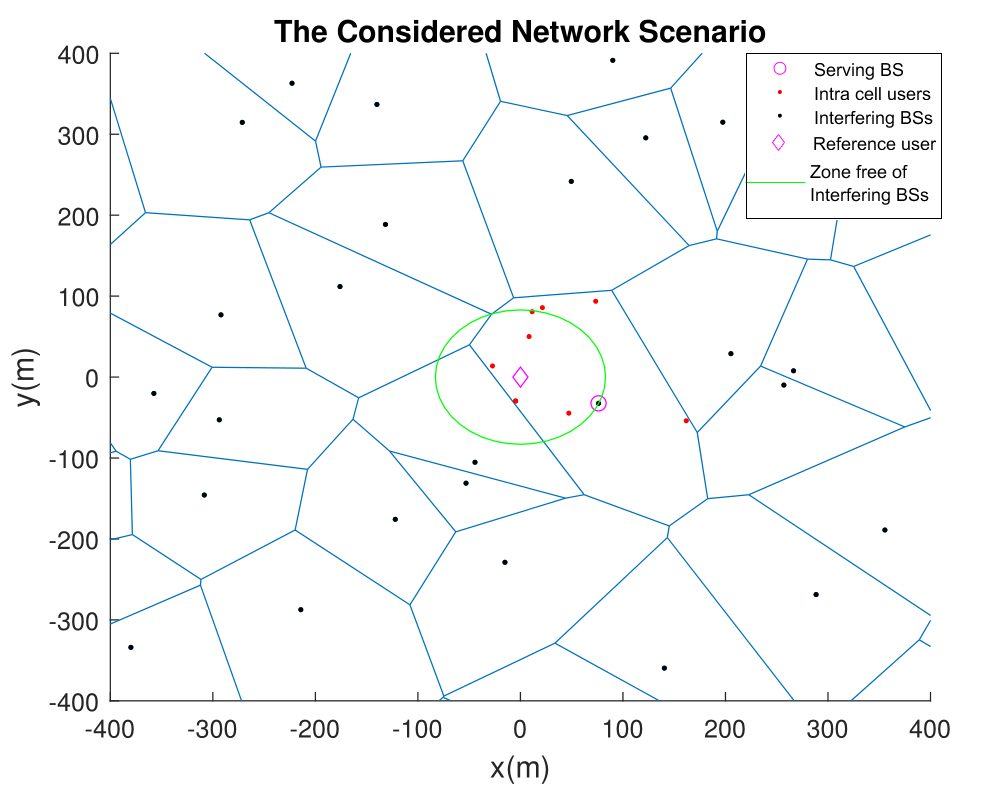}}
    \caption{The considered network scenario.}
    \label{fig:Scenario}
  \end{center}
\end{figure}

The considered scenario is depicted in Figure \ref{fig:Scenario}. ${U\!E}_0$ is marked by the magenta diamond, $BS_0$ by the magenta circle and the intra cell users are depicted in red. The interfering BSs are depicted in black and reside in distances ($||x_i||>||x_0||$), where the circle of radius $||x_0||$ around ${U\!E}_0$  is depicted in green. Every BS residing in the region outside this circle is acting as an interferer in the DL.

\section{MGF of the Aggregate Other-cell Interference}
\label{sec:MGF}
Having presented the considered network scenario and having defined the set of interfering BSs by means of their distance to the origin, we can proceed with the mathematical formulation of the aggregate other-cell interference and the MGF of the latter.
\subsection{Derivation of the MGF}
\label{ssec:MGF_Der}

The aggregate other-cell interference in the DL is mathematically formulated as follows:
\begin{align}
\label{eqn:InterferenceDef}
g=\sum\limits_{\substack{x \in \Psi, \\ x \neq x_0}} \frac{P_{tx}}{\kappa \left\lVert x \right\rVert^{\beta}},
\end{align}  
where 
$P_{tx}$ denotes the transmit power of the BSs, $\beta$ the path-loss exponent, $\left\lVert x \right\rVert$ the distance of the interferer to the origin and $\kappa$ the path-loss at a reference distance of 1 meter. 

Although a term of fast fading, denoted by ${\left| {{h^{\left( x \right)}}} \right|}^2$, is present at the propagation, it has not been introduced in the previous expression (\ref{eqn:InterferenceDef}). The reason for that is that taking into account the fast fading of the interferers during the computation of the UE capacity would implicitly mean that the UE has perfect knowledge of the channel of all interferers. However, since this is not the case in practice, averaging over this fading would provide an upper bound for the capacity. As opposed to that, the omission of the fast fading ${\left| {{h^{\left( x \right)}}} \right|}^2$ provides a lower bound for the capacity \cite{Geordie}. Since a worst case scenario analysis is more sensible than an overoptimistic calculation of the achievable rate, the fast fading of the interferers is not introduced. However, if accounting for the fast fading is of interest, this can be done without increasing significantly the complexity of the presented analysis as will be demonstrated in Section \ref{ssec:Fading}.

The set of interfering BSs in (\ref{eqn:InterferenceDef}) has been defined by means of their distances to the origin, i.e. $\left\lVert x_i \right\rVert>\left\lVert x_0 \right\rVert$. Equivalently, by defining the path-loss as follows:
\begin{align}
\label{eqn:DefPathLoss}
L^{(x)}=\kappa \left\lVert x \right\rVert^\beta,
\end{align}  
the set of interfering BSs can be defined by means of their path-loss to the origin (i.e. $ {{L^{\left( x \right)}} > {L^{\left( 0 \right)}}}$, where ${{L^{\left( 0 \right)}}}={{L^{\left( x_0 \right)}}}$ for brevity in the notation). Having defined mathematically the set of interfering BSs, the MGF of $g$ can be obtained by employing the probability generating fuctional (PGFL) theorem according to which \cite{Haenggi}:
\begin{align}
\label{eqn:PGFL}
\mathbb{E}\left\{\prod_{x \in \Psi}^{} f(x)\right\}&  \mathop = \limits^{\left( a \right)}  \exp \left(-\lambda  \int_{\mathbb{R}^2}^{} (1-f(x))~dx \right) \nonumber\\
& \mathop = \limits^{\left( b \right)}\exp \left( \int_{\mathbb{R}}^{} (f(r)-1) ~2 \pi \lambda r dr \right) \nonumber \\
& \mathop = \limits^{\left( c \right)}\exp \left( \int_{\mathbb{R}}^{} (f(r)-1) ~\frac{2 \pi \lambda}{\beta}\left(\frac{1}{\kappa}\right)^{\frac{2}{\beta}} y^{\frac{2}{\beta}-1} dy \right) ,
\end{align}  
where (\ref{eqn:PGFL}b) is obtained by computing the double integral of (\ref{eqn:PGFL}a) in polar coordinates for $0 \leq \theta \leq 2 \pi$ and for $r=\left\lVert x \right\rVert$, and (\ref{eqn:PGFL}c) is obtained by changing the variable of the integration to the path-loss according to (\ref{eqn:DefPathLoss}).
Hence, employing (\ref{eqn:PGFL}c) the MGF of $g$ is given by \cite{Thanh}:

\begin{align}
\label{eqn:MGFinter}
	M_{g}\left( s; L^{\left( 0 \right)} \right)&={ \mathbb{E}_{\Psi }}\left\{ \exp \left( - s g \right) \right\}
	\nonumber \\
& ={ \mathbb{E}_{\Psi }}\left\{ {\exp \left( { - s \sum\limits_{x \in \Psi} \frac{{{P_{\rm{tx}}}}}{{{L^{\left( x \right)}}}}~ {\bf{1}} \left( {{L^{\left( x \right)}} > {L^{\left( 0 \right)}}} \right)} \right)} \right\}
	\nonumber \\
	&={ \mathbb{E}_{\Psi }}\left\{ {\prod\limits_{x \in \Psi } {} \exp \left( { - s\frac{{{P_{\rm{tx}}}}}{{{L^{\left( x \right)}}}}~ {\bf{1}} \left( {{L^{\left( x \right)}} > {L^{\left( 0 \right)}}} \right)} \right)} \right\}
	\nonumber \\	
	 &\mathop = \limits^{\left( a \right)} \exp \left(  \! \; {\int\limits_{{L^{\left( 0 \right)}} }^\infty {} \! \! \! \! \! \! \left( {\exp \left( { - s\frac{{{P_{\rm{tx}}}}}{y}} \right) - 1} \right) {\frac{{2\pi \lambda }}{\beta }\left( \frac{1}{\kappa}\right)^{\frac{2}{\beta}}{y^{\frac{2}{\beta } - 1}}} dy} \right)
	\nonumber \\
	 & \mathop = \limits^{\left( b \right)} \exp { \!\! {} \left( \! {\pi \lambda {{\left( {\frac{{ L^{\left( 0 \right)} }}{\kappa }} \right)}^{\frac{2}{\beta }}} \! \left( \! {1 - \! {\;_1}{F_1}\left( { - \frac{2}{\beta },1 - \frac{2}{\beta },\frac{{ - s{P_{\rm{tx}}}}}{{ L^{\left( 0 \right)} }}} \! \right)} \! \right)} \! \right), \!} \! \nonumber \\& s \in \mathbb{R},
\end{align}
where the second argument of $M_{g}\left( s; L^{\left( 0 \right)} \right)$ denotes the dependence of the MGF on the random variable $L^{\left( 0 \right)}$ and ${\bf{1}}(\cdot)$ is the indicator function. (a) holds by employing (\ref{eqn:PGFL}c), and (b) is attained by using the result of \cite{HyperGeometric} according to which:
\begin{align}
\label{eqn:HyperGeometric}
I =& \int\limits_{a}^\infty \!\! {} \left( \exp \left( \frac{b}{z}\right) -1\right) z^{v-1}dz \nonumber \\
=&\frac{1}{v} a^v \left(1-{\;_1}{F_1}\left(-v,1-v,\frac{b}{a} \right) \right),
\end{align}
with $\! {\;_1}{F_1}(a,b,z)$ being the Kummer confluent hypergeometric function given by:
\begin{align}
\label{eqn:HyperGeometricDef}
\! {\;_1}{F_1}(a,b,z) = \sum\limits_{k=0}^\infty \frac{(a)_k z^k}{(b)_k k!},
\end{align}
where $(\cdot)_k$ denotes the Pochhammer function given by:
\begin{align}
\label{eqn:RisingFactorial}
(x)_k = \prod_{n=0}^{k-1}(x+n).
\end{align}

The definitions of (\ref{eqn:HyperGeometricDef}) and (\ref{eqn:RisingFactorial}) demonstrate the intractability of (\ref{eqn:MGFinter}b). Since the derivation of the MGF of the aggregate interference is one of the fundamental tools of stochastic geometry, the intractability of (\ref{eqn:MGFinter}b) propagates to every analysis employing the MGF, hindering the derivation of closed form figures of merit.

\subsection{MGF Approximation}
\label{ssec:MGF_App}
In order to overcome the limitations imposed by (\ref{eqn:MGFinter}b), the present paper introduces a simple, albeit extremely accurate, approximation of the MGF by introducing an alternative calculation of (\ref{eqn:MGFinter}a). In this course, the MGF is derived as follows:

\begin{align}
\label{eqn:MGF_Approx}
{M_{g}}\left( s; { L^{\left( 0 \right)} } \right) & \mathop = \limits^{\left( a \right)}
 \exp \left( { \; {\int\limits_{{L^{\left( 0 \right)}} }^\infty {} \left( {\exp \left( { - s\frac{{{P_{\rm{tx}}}}}{y}} \right) - 1} \right) {\frac{{2\pi \lambda }}{\beta }\left( {\frac{1}{\kappa }} \right)^{\frac{2}{\beta}}{y^{\frac{2}{\beta } - 1}}} dy}} \right)\nonumber \\
 &\mathop = \limits^{\left( b \right)} \exp \left( { \; {\int\limits_{ {L^{\left( 0 \right)}} }^\infty {} {\left(\sum \limits_{n=1}^\infty \frac{1}{n!}\left(- s \frac{{{P_{\rm{tx}}}}}{y} \right)^n \right) } {\frac{{2\pi \lambda }}{\beta }\left( {\frac{1}{\kappa }} \right)^{\frac{2}{\beta}}{y^{\frac{2}{\beta } - 1}}} dy}} \right) \nonumber \\
&\mathop = \limits^{\left( c \right)}\exp \left( \frac{2 \pi \lambda}{\beta}  \left(\frac{L^{\left( 0 \right)}}{\kappa} \right)^\frac{2}{\beta} \left(\sum \limits_{n=1}^\infty \frac{(-1)^{n+1} ( s P_{\rm{tx}} )^n}{ \left({L^{\left( 0 \right)}} \right)^n n! (\frac{2}{\beta}-n)} \right) \right) \nonumber \\
  &\mathop = \limits^{\left( d \right)}  \exp \left(\substack{-\pi \lambda \left(\frac{L^{\left( 0 \right)}}{\kappa} \right)^\frac{2}{\beta}  \left( \exp\left(\frac{-s P_{\rm{tx}}}{{L^{\left( 0 \right)}}} \right)-1+\left(\frac{s P_{\rm{tx}}}{L^{\left( 0 \right)}}\right)^{\frac{2}{\beta}}\Gamma\left(1-\frac{2}{\beta},0,\frac{s P_{\rm{tx}}}{L^{\left( 0 \right)}}\right) \right)} \right)  \nonumber \\ 
&\mathop = \limits^{\left( e \right)} \exp \left(\substack{-\pi \lambda \left(\frac{L^{\left( 0 \right)}}{\kappa} \right)^\frac{2}{\beta} \left(  \exp\left(\frac{-s P_{\rm{tx}}}{{L^{\left( 0 \right)}}} \right)-1+\left( \frac{s P_{\rm{tx}}}{L^{\left( 0 \right)}}\right)^{\frac{2}{\beta}} \int\limits_{0}^{\frac{s P_{\rm{tx}}}{L^{\left( 0 \right)}}}t^{-\frac{2}{\beta}}\exp(-t) dt\right)}\right),
\end{align}  
where (b) holds by employing the Taylor expansion of the exponential term, (c) holds by a simple calculation of the integral, and (d) and (e) are obtained from the definition of the generalized incomplete gamma function $\Gamma(\cdot,\cdot,\cdot)$.  

Having defined (\ref{eqn:MGF_Approx}e) it can be noted that the term within the integral of (\ref{eqn:MGF_Approx}e) eventually tends to 0. When this happens, namely when $ \exp\left(\frac{-s P_{\rm{tx}}}{{L^{\left( 0 \right)}}} \right) \approx 0$, the integral converges to a constant value. Hence, (\ref{eqn:MGF_Approx}e) can be approximated by a piecewise function, involving a constant value when $ \exp\left(\frac{-s P_{\rm{tx}}}{{L^{\left( 0 \right)}}} \right) \approx 0$ and a varying function when $ \exp\left(\frac{-s P_{\rm{tx}}}{{L^{\left( 0 \right)}}} \right) \neq 0$.

In particular, by employing the constant value to which the integral converges when $ \exp\left(\frac{-s P_{\rm{tx}}}{{L^{\left( 0 \right)}}} \right) \approx 0$, it holds that:

\begin{align}
\label{eqn:Appr_Sec}
&\exp\left(\frac{-s P_{\rm{tx}}}{{L^{\left( 0 \right)}}} \right)-1+\left( \frac{s P_{\rm{tx}}}{L^{\left( 0 \right)}}\right)^{\frac{2}{\beta}} \int\limits_{0}^{\frac{s P_{\rm{tx}}}{L^{\left( 0 \right)}}}t^{-\frac{2}{\beta}}\exp(-t) dt \nonumber \\  
& \mathop  \approx \limits^{\left(  \exp\left(\frac{-s P_{\rm{tx}}}{{L^{\left( 0 \right)}}} \right) \approx 0 \right)} \left( \frac{s P_{\rm{tx}}}{L^{\left( 0 \right)}}\right)^{\frac{2}{\beta}} \Gamma\left(1-\frac{2}{\beta}\right) -1
\end{align}
and when $ \exp\left(\frac{-s P_{\rm{tx}}}{{L^{\left( 0 \right)}}} \right) \neq 0$, (\ref{eqn:MGF_Approx}e) can be approximated by the Taylor expansion around 0 as follows:
\begin{align}
\label{eqn:Appr_Fir}
\exp\left(\frac{-s P_{\rm{tx}}}{{L^{\left( 0 \right)}}} \right)-1+\left( \frac{s P_{\rm{tx}}}{L^{\left( 0 \right)}}\right)^{\frac{2}{\beta}} \int\limits_{0}^{\frac{s P_{\rm{tx}}}{L^{\left( 0 \right)}}}t^{-\frac{2}{\beta}}\exp(-t) dt = \sum \limits_{n=1}^\infty \frac{-2 (- s P_{\rm{tx}} )^n}{ \left({L^{\left( 0 \right)}} \right)^n n! (n \beta -2)}.
\end{align}

Thus, by combining (\ref{eqn:Appr_Sec}) and (\ref{eqn:Appr_Fir}), (\ref{eqn:MGF_Approx}e) can be approximated as follows:
\begin{align}
\label{eqn:MGF_Approx_Full}
&{M_{g}}\left( s; { L^{\left( 0 \right)} } \right) \approx 
   &\begin{cases}
       \exp \left( \pi \lambda \left(\frac{L^{\left( 0 \right)}}{\kappa} \right)^\frac{2}{\beta} \sum \limits_{n=1}^\infty \frac{2 (- s P_{\rm{tx}} )^n}{ \left({L^{\left( 0 \right)}} \right)^n n! (n \beta -2)} \right),&\!\!\! \frac{s P_{\rm{tx}}}{L^{\left( 0 \right)}}\leq c,\\
       \exp \left( \pi \lambda \left(\frac{L^{\left( 0 \right)}}{\kappa} \right)^\frac{2}{\beta}\left(-\left( \frac{s P_{\rm{tx}}}{L^{\left( 0 \right)}}\right)^{\frac{2}{\beta}} \Gamma\left(1-\frac{2}{\beta}\right) +1 \right) \right),&\!\!\!  \frac{s P_{\rm{tx}}}{L^{\left( 0 \right)}}> c
    \end{cases}
\end{align} 
and by employing only the first two terms of the Taylor expansion, (\ref{eqn:MGF_Approx_Full}) can be approximated by:
\begin{align}
\label{eqn:MGF_Approx_Short}
&{M_{g}}\left( s; { L^{\left( 0 \right)} } \right) \! \approx 
    & \begin{cases}
           \exp \left( \pi \lambda \left(\frac{L^{\left( 0 \right)}}{\kappa} \right)^\frac{2}{\beta} \left(\frac{-2 }{(\beta-2)}\frac{s P_{\rm{tx}}}{L^{\left( 0 \right)}}+\frac{1}{2\beta-2}\left(\frac{s P_{\rm{tx}}}{L^{\left( 0 \right)}}\right)^2 \right) \right)\!\!,& \!\!\!\frac{s P_{\rm{tx}}}{L^{\left( 0 \right)}}\leq c\\
          \exp \left( \pi \lambda \left(\frac{L^{\left( 0 \right)}}{\kappa} \right)^\frac{2}{\beta}\left(-\left( \frac{s P_{\rm{tx}}}{L^{\left( 0 \right)}}\right)^{\frac{2}{\beta}} \Gamma\left(1-\frac{2}{\beta}\right) +1 \right) \right)\!\!,&\!\!\! \frac{s P_{\rm{tx}}}{L^{\left( 0 \right)}}> c
         \end{cases}
\end{align}  
where $c$ is a constant indicating the point after which the integral of (\ref{eqn:MGF_Approx}e) converges to a constant value. $c$ is the point of intersection of the two functions of (\ref{eqn:MGF_Approx_Short}) and can be obtained by solving the following equation:
\begin{align}
\label{eqn:c}
\frac{-2c}{\beta-2}+\frac{c^2}{2\beta-2}=-c^{\frac{2}{\beta}}\Gamma\left(1-\frac{2}{\beta}\right)+1.
\end{align}  
Although, (\ref{eqn:c}) cannot be solved analytically, for the limited range of the path loss exponent $\beta$ (i.e. $\beta \in [2,5]$) it can be computed numerically and the value of $c$ for any value of $\beta \in [2,5]$ can be approximated by:
\begin{align}
\label{eqn:c_eq}
c \approx 0.06662 \log(\beta-1.528)+1.227,
\end{align}  
where $\log(\cdot)$ in (\ref{eqn:c_eq}) and henceforth denotes the natural logarithm.

The employment of the Taylor expansion and of the exact value of the generalized incomplete function $\Gamma(\cdot,\cdot,\cdot)$ at the extreme cases guarantees the tightness of the approximation far from the intersection point $c$. However, the tightness of the proposed approximation still needs to be verified close to $c$. In this course, relevant figures will be provided in the following sections validating the tightness of the approximation close to $c$ against the exact result of (\ref{eqn:MGFinter}b). Given the dependence of $c$ on $\beta$, the provided figures will demonstrate the tightness of the proposed approximation in the whole range of $\beta \in [2,5]$, which is also the range of interest in wireless networks. It should also be noted, that the value of $\lambda$ does not have any impact on the tightness of the approximation, since the term of (\ref{eqn:MGF_Approx}e) that has been approximated by (\ref{eqn:Appr_Sec}) and (\ref{eqn:Appr_Fir}) does not involve $\lambda$. Therefore, the proposed approximation of (\ref{eqn:MGF_Approx_Short}) is not just far more tractable and simple than (\ref{eqn:MGFinter}b), but is also tight over the whole range of values that are of interest in wireless networks. 

\subsection{Fast Fading of Interferers}
\label{ssec:Fading}
The aforementioned analysis can be easily extended to account also for the fast fading of the interferers, if the latter is of interest. In this course, an independently marked point process can be employed, that is, a point process where a random variable, known as mark and denoted by $M_x$, is randomly assigned to each random point of the point process $x$ \cite{Haenggi}. The marks are mutually independent and the conditional distribution of mark $M_x \in \mathbb{R}^l$ of a point $x \in \Psi$ depends only on the location of $x$. For an independently marked homogeneous PPP with density $\lambda$ on $ \mathbb{R}^2$ and marks with distribution $F_x (dM)$ on $ \mathbb{R}^l$, the Laplace transform of a function $f(x,M_x)$ is given by \cite{MarkedPPP}:
\begin{align}
\label{eqn:Laplace}
\!\!\!\!\!\!\!\!\!\!\!\!\!\mathcal{L}_\Psi(f) &=\mathbb{E}\left\{\exp(- \sum\limits_{x \in \Psi} f(x,M_x))\right\}\nonumber \\
\!\!\!\!\!\!\!\!\!\!\!\!\!\!\!\!&= \exp\left({ \!\!\!-\lambda  \!\!\int_{ \mathbb{R}^2}^{}\!\!\left(\!1-\!\!\!\int_{ \mathbb{R}^l}^{} \!\!\exp\left(-f(x,M_x)\right)F_x(dM)\right) }\!\!\right).
\end{align}  

Hence, if the fast fading of the interferes needs be introduced in (\ref{eqn:InterferenceDef}), (\ref{eqn:Laplace}) can be applied directly to (\ref{eqn:MGFinter}), with $M_x=\left| {{h^{\left( x \right)}}} \right| \in \mathbb{R}$, in order to compute the expectation with respect to the path-loss and to the fading of the interferers. That is, by setting: 
\begin{align}
\label{eqn:Laplace_Fading}
f(x,h)={ s \frac{{{P_{\rm{tx}}{\left| {{h^{\left( x \right)}}} \right|}^2}}}{{{L^{\left( x \right)}}}}~ {\bf{1}} \left( {{L^{\left( x \right)}} > {L^{\left( 0 \right)}}} \right)}
\end{align}  
and $l=1$ and by employing the distribution $F_H(h)$ for the respective type of fading the MGF of  (\ref{eqn:MGFinter}a) is revised as follows: 
\begin{align}
\label{eqn:MGFinterFading}
	&\!\!\!\!\!\!\!\!\!\!\!\!M_{g}\left( s; L^{\left( 0 \right)} \right)= \exp \left(  \! \; {\int\limits_{h=0}^\infty \left[\int\limits_{{L^{\left( 0 \right)}} }^\infty {} \left( {\exp \left( { - s\frac{{{P_{\rm{tx}}{\left| {h} \right|}^2}}}{y}} \right) - 1} \right) {\frac{{2\pi \lambda }}{\beta }\left( \frac{1}{\kappa}\right)^{\frac{2}{\beta}}{y^{\frac{2}{\beta } - 1}}} dy\right]F_H(h)} \right), \! \nonumber \\
&\!\!\!\!\!\!\!\!\!\!\!\! s \in \mathbb{R}.
\end{align}
The employment of (\ref{eqn:Laplace}) allowed for moving the expectation over the fading within the exponential term of (\ref{eqn:MGFinterFading}), thus, simplifying the analysis to a great extent. As a result, the expectation over the fading of the interferes can also be moved within the exponential term of (\ref{eqn:MGF_Approx_Short}). Given the tractability of (\ref{eqn:MGF_Approx_Short}), the introduction of an additional integral within the exponential term has a minor effect on the complexity of the derived expressions and the analysis can be easily extended accordingly.

However, as already mentioned, the present analysis does not account for the fading of the interferers focusing on a realistic scenario where the UE does not have perfect knowledge of the channel of all interferers.

\section{Ergodic Capacity in the DL}
\label{sec:Ergodic}
Having defined a simple approximation of the MGF, the latter can be employed to provide closed form expressions for the DL ergodic rate for the interference limited case. In this course, the analysis will commence by employing the MGF of (\ref{eqn:MGFinter}b) and (\ref{eqn:MGF_Approx_Short}) to derive the coverage probability for the interference limited case. The latter can be derived by both expressions, i.e. (\ref{eqn:MGFinter}b) and (\ref{eqn:MGF_Approx_Short}), in spite of the intractability of the former, allowing for the comparison of the two results and, thus demonstrating the accuracy of the introduced approximation. Subsequently, capitalizing on the accuracy of the introduced approximation, the DL ergodic rate will be derived in closed form by employing the introduced approximate expression for the coverage probability. 

The analysis which will initially account only for networks comprising much more users than BSs will then be extended to networks comprising more BSs than users.

\subsection{Probability of Coverage}
\label{ssec:Pcov}

The probability of coverage (i.e. the probability of the SINR exceeding the value $\gamma$) is defined as follows:
\begin{align}
\label{eqn:Pcov}
P_{cov}&= \mathbb{P}\left(\frac{P_{\rm{tx}} \left|h^{(0)} \right|^2/L^{\left( 0 \right)}}{\sum \limits_{x \in \Psi, x \neq x_0}P_{\rm{tx}}/L^{\left( i \right)}+\sigma_N^2} \geq \gamma \right)\\ 
& =  \mathbb{P}\left( \left|h^{(0)} \right|^2 \geq \frac{\gamma L^{\left( 0 \right)} \left(g+\sigma_N^2 \right) }{P_{\rm{tx}}} \right),
\end{align}
where $\sigma_N^2$ denotes the noise power and $h^{(0)}$ the fast fading of the intended user. As opposed to the fast fading of the interferers, the fast fading of the intended user is estimated and known in practice.  

Assuming Rayleigh fading, then the random variable $\left|h^{(0)} \right|^2 $ follows an exponential distribution with unit mean and the $P_{cov}$ is given by:
\begin{align}
\label{eqn:Pcov1}
\!\!\!\!\!\!\!\!\!\!\!P_{cov}& \mathop = \limits^{\left( a \right)} \mathbb{E}_{g,L^{\left( 0 \right)}}\left\{\exp\left(-\frac{\gamma L^{\left( 0 \right)} (g+\sigma_N^2)}{P_{\rm{tx}}}\right)\right\}\nonumber \\
\!\!\!\!\!\!\!\!\!\!\!&  \mathop = \limits^{\left( b \right)}  \mathbb{E}_{L^{\left( 0 \right)}}\left\{\exp\left(-\frac{\gamma L^{\left( 0 \right)} \sigma_N^2}{P_{\rm{tx}}}\right) M_{g}\left(\frac{\gamma L^{\left( 0 \right)}}{P_{\rm{tx}}}; L^{\left( 0 \right)}\right) \right\} \nonumber \\
\!\!\!\!\!\!\!\!\!\!\!&  \mathop = \limits^{\left( c \right)} \int \limits_{y=0}^\infty \exp \left(-\frac{ \gamma y \sigma_N^2}{P_{\rm{tx}}} \right) {M_{g}}\left( \frac{\gamma y}{P_{\rm{tx}}}; y \right) f_{L^{\left( 0 \right)}}(y) 
dy,
\end{align}
where (a) is obtained by the CCDF of the exponential distribution, (b) is obtained based on the definition of the MGF in (\ref{eqn:MGFinter}), and (c) from computing the expectation with respect to the path-loss $L^{\left( 0 \right)}$ to the serving BS $BS_0$. 

The probability density function (PDF) of the distance between a reference user and its closest BS for a PPP is given in \cite{PDF}. By employing this PDF and the definition of (\ref{eqn:DefPathLoss}), the PDF of the path-loss between a reference user and its closest BS $f_{L^{\left( 0 \right)}}(y)$ is given by:
\begin{align}
\label{eqn:PDF}
f_{L^{\left( 0 \right)}}(y)=\frac{2\pi\lambda}{\beta}\left(\frac{1}{\kappa}\right)^{\frac{2}{\beta}}y^{\frac{2}{\beta}-1}\exp\left(-\pi\lambda \left(\frac{y}{\kappa}\right)^{\frac{2}{\beta}}\right),
\end{align}

In the interference limited case, the exponential term of (\ref{eqn:Pcov1}c) is equal to 1 and, by employing (\ref{eqn:PDF}), the coverage probability can be calculated as follows:
\begin{align}
\label{eqn:Pcov2}
\!\!\!\!\!\!\!\!\!\!\!\!\!\!P_{cov}& = \int \limits_{0}^\infty {M_{g}}\left( \frac{y \gamma}{P_{\rm{tx}}}; y \right)\!\frac{2\pi\lambda}{\beta}\left(\frac{1}{\kappa}\right)^{\frac{2}{\beta}}\!\!y^{\frac{2}{\beta}-1}\exp\left(\!\!\!-\pi\lambda \left(\frac{y}{\kappa}\right)^{\frac{2}{\beta}}\!\!\right) dy\nonumber \\
\!\!\!\!\!\!\!\!\!\!\!\!\!\!& \mathop = \limits^{\left( a \right)}\frac{1}{ \! {\;_1}{F_1}\left(-\frac{2}{\beta},1-\frac{2}{\beta},-\gamma\right)}\nonumber \\
\!\!\!\!\!\!\!\!\!\!\!\!\!\!& \mathop \approx \limits^{\left( b \right)}\frac{{\bf{1}} \left( \gamma \leq c \right)}{\left(-\frac{\gamma^2}{2\beta-2}+\frac{2\gamma}{\beta-2}+1\right)} +\frac{{\bf{1}} \left( \gamma>c \right)}{\gamma^{\frac{2}{\beta}}\Gamma(1-\frac{2}{\beta})},
\end{align}
where (a) is obtained by employing (\ref{eqn:MGFinter}b) and (b) by employing the approximate MGF of (\ref{eqn:MGF_Approx_Short}). 

\begin{figure}[h!]
  \begin{center}
   \centerline{
    \includegraphics[height=9cm]{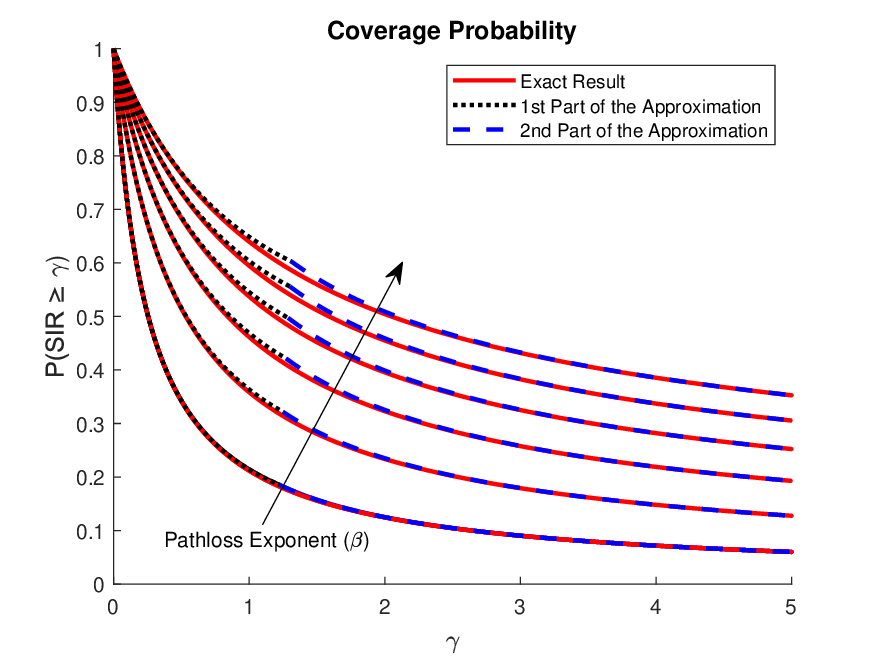}}
    \caption{Probability of coverage for different path loss exponent values $2.5\leq \beta \leq 5$. $\beta$ increases in the direction of the arrow with a step of 0.5.}
    \label{fig:Pcov}
  \end{center}
\end{figure}

The results of (\ref{eqn:Pcov2}) verify the theoretical results presented in Section \ref{sec:introduction}, since for a single-slope path loss model and for more users than BSs, the coverage probability of (\ref{eqn:Pcov2}) does not depend on the density of BSs $\lambda$, but only on the path loss exponent $\beta$ and the SIR value $\gamma$ ($c$ is also a function of $\beta$ given by (\ref{eqn:c_eq})). The accuracy of (\ref{eqn:Pcov2}b) and, implicitly, the accuracy of the MGF of (\ref{eqn:MGF_Approx_Short}), is demonstrated in Fig. \ref{fig:Pcov} where the approximate coverage probability of (\ref{eqn:Pcov2}b) is compared against the exact result of (\ref{eqn:Pcov2}a). Apart from extremely accurate, the approximate coverage probability of (\ref{eqn:Pcov2}b) is also significantly more tractable than the exact result of (\ref{eqn:Pcov2}a).

\subsection{Ergodic Rate}
\label{ssec:Rate}

Having defined the tractable and accurate approximation of (\ref{eqn:Pcov2}b) for the probability of coverage, this can be employed to compute the DL ergodic rate. In particular, the probability of coverage given by (\ref{eqn:Pcov2}b) constitutes the CCDF of the SIR (i.e. for $SIR=w$, $P_{cov}=1-F_W(w)$). Hence, the derived tractable expression of the CCDF of the SIR allows for the computation of the DL ergodic rate by averaging over the SIR as follows:
\begin{align}
\label{eqn:Rate}
R &= \mathbb{E}_w\{\log(1+w)\}= \int \limits_{0}^\infty \log(1+w) F_W'(w) dw\nonumber \\
&\mathop = \limits^{\left( a \right)} \int \limits_{0}^\infty \frac{P_{cov}}{1+w} dw \nonumber \\
&\mathop \approx \limits^{\left( b \right)} \int \limits_{0}^c \frac{1}{\left(-\frac{w^2}{2\beta-2}+ \frac{2 w}{\beta-2}+1 \right)(1+w)} dw + \int \limits_{c}^\infty \frac{1}{\left(w^{\frac{2}{\beta}} \Gamma\left(1-\frac{2}{\beta}\right) \right)(1+w)} dw,
\end{align}
where (a) is obtained by integrating by parts and (b) is obtained by employing (\ref{eqn:Pcov2}b). 
The employment of (\ref{eqn:Pcov2}a) in (\ref{eqn:Rate}a) would not allow the analytical computation of the above integral. However, (\ref{eqn:Rate}b) can be computed in closed form and the rate is given by:
\begin{align}
\label{eqn:Rate_Closed}
R =& (2\beta-2)\Bigg(\!\frac{4+2\alpha-3\beta-\alpha \beta}{2\alpha(10-11\beta+2\beta^2)}\bigg(\!\!\log{\left(\frac{c+\alpha-\frac{2\beta-2}{\beta-2}}{\alpha-\frac{2\beta-2}{\beta-2}}\right)}\!\!\bigg) \nonumber \\
&+ \frac{-4+2\alpha+3\beta-\alpha \beta}{2\alpha(10-11\beta+2\beta^2)}\bigg(\log{\left(\frac{c-\alpha-\frac{2\beta-2}{\beta-2}}{-\alpha-\frac{2\beta-2}{\beta-2}}\right)}\bigg) \nonumber \\
&+\frac{-2+\beta}{(10-11\beta+2\beta^2)}\bigg(\log{\left(c+1\right)}\bigg) \Bigg) \nonumber \\
&+ \frac{\beta c^{-\frac{2}{\beta}}}{2\Gamma\left(1-\frac{2}{\beta}\right)} \! {\;_2}{F_1}\left(1,\frac{2}{\beta},\frac{2+\beta}{\beta},-\frac{1}{c}\right),
\end{align}
where the first three terms of (\ref{eqn:Rate_Closed}) are obtained by the calculation of the first term of (\ref{eqn:Rate}b) and the last term of (\ref{eqn:Rate_Closed}) is obtained by the calculation of the last term of (\ref{eqn:Rate}b). $\! {\;_2}{F_1}(a,b,c,z)$ denotes the Gaussian hypergeometric function, $\alpha=\sqrt{\left(\frac{2\beta-2}{\beta-2}\right)^2+2\beta-2}$ and $c$ is given by (\ref{eqn:c_eq}). Again, as in the case of  (\ref{eqn:Pcov2}), (\ref{eqn:Rate_Closed}) depends only on the path loss exponent $\beta$.

\begin{figure}[h!]
  \begin{center}
   \centerline{
    \includegraphics[height=9cm]{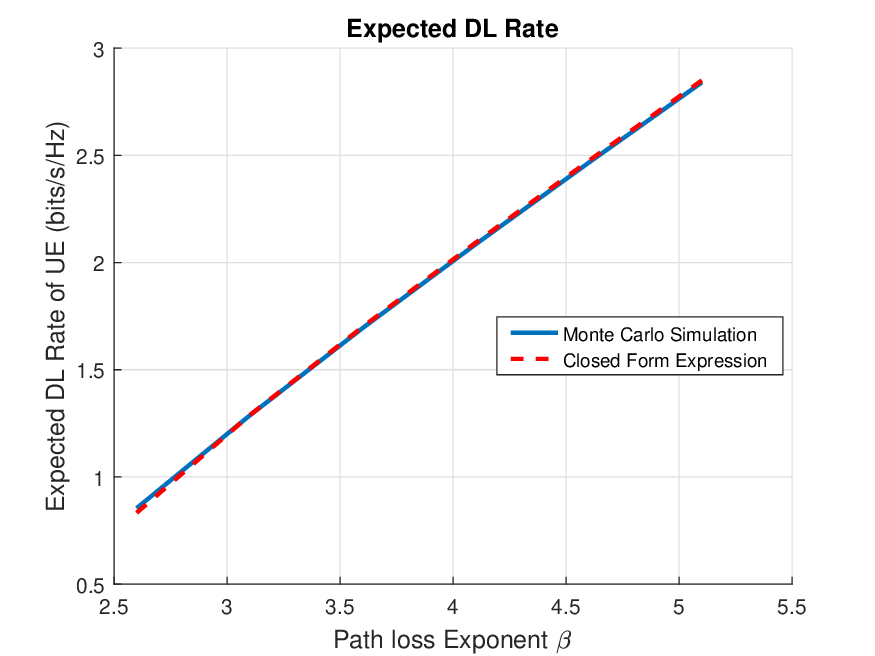}}
    \caption{DL ergodic capacity vs path loss exponent for the interference limited case.}
    \label{fig:Rate}
  \end{center}
\end{figure}

The employment of (\ref{eqn:Rate_Closed}) allows the computation of the rate for the interference limited case in closed form. The tight performance of (\ref{eqn:Rate_Closed}) for the calculation of the ergodic rate is demonstrated in Fig. \ref{fig:Rate}, where (\ref{eqn:Rate_Closed}) is compared against the results obtained by Monte Carlo simulations. Even though the rate does not depend on the density of the BSs, the density employed for generating the simulation of Fig.  \ref{fig:Rate} was $\lambda = 1.27e-06$.

Several research works have focused on deriving expressions for the DL ergodic rate, since the latter constitutes the most sensible figure of merit for evaluating the performance of UDNs. Indicatively, in \cite{Marco2} the authors have provided expressions for the DL ergodic rate in heterogeneous cellular networks for all different types of fading. However, in all of these works, including the latter, the calculation of the DL ergodic rate involved at least one integration that had to be computed numerically. This imposed inherent limitations to the applicability of those expressions to complex optimization problems. In order to overcome this problem the authors of \cite{Lozano} provided for the first time in the literature expressions for the DL ergodic rate in the interference limited case, for a fully loaded scenario (i.e. for $\lambda_{UE}\gg \lambda$) that did not involve any numerical integration. In this direction, the authors of \cite{Lozano} provided lookup tables and employed the Meijer-G function to provide a tight approximation of the DL ergodic rate in the fully loaded case. In particular, the approximate ergodic rate of \cite{Lozano} is given by:
\DeclarePairedDelimiterX\MeijerM[3]{\lparen}{\rparen}%
{\begin{smallmatrix}#1 \\ #2\end{smallmatrix}\delimsize\vert\,#3}

\newcommand\MeijerG[8][]{%
  G^{\,#2,#3}_{#4,#5}\MeijerM[#1]{#6}{#7}{#8}}

\begin{figure}[h!]
  \begin{center}
   \centerline{
    \includegraphics[height=9cm]{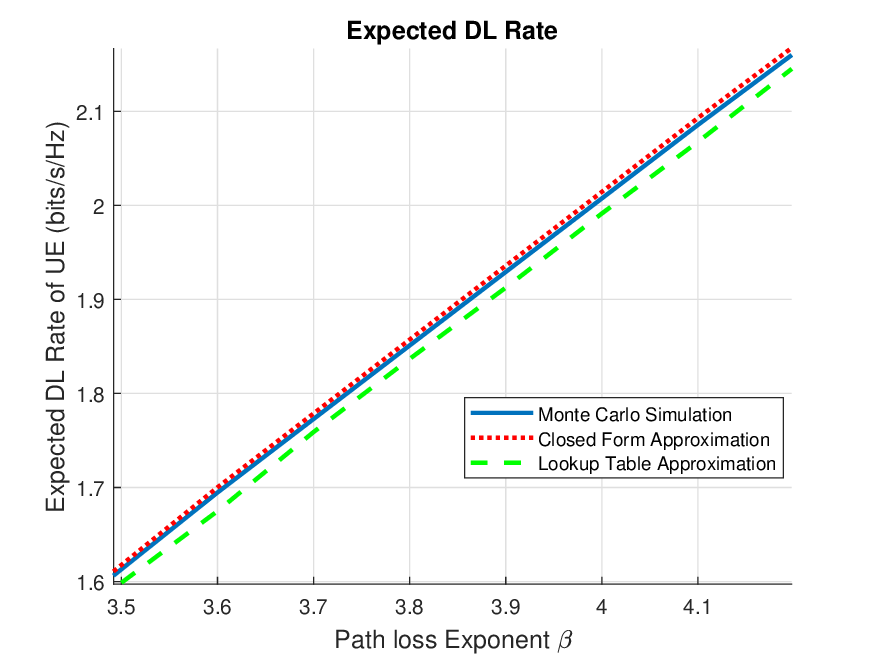}}
    \caption{DL ergodic capacity vs path loss exponent for the interference limited case. Tightness of closed form approximation compared to the lookup table approximation employing the Meijer-G function. The range of $\beta$ is defined by the range of the respective lookup table provided in \cite{Lozano}.}
    \label{fig:RateLoz}
  \end{center}
\end{figure}

\WithSuffix\newcommand\MeijerG*[7]{%
  G^{\,#1,#2}_{#3,#4}\MeijerM*{#5}{#6}{#7}}
\begin{align}
\label{eqn:Rate_Lozano}
C =& \frac{-s^* \log_2 e}{1+s^*}\left(\operatorname{E_1}\left(\frac{-s^*}{D_\delta}\right)-\exp\left(\frac{1+s^*}{D_\delta}\right)\operatorname{E_1}\left(\frac{1}{D_\delta}\right)\right) \nonumber \\&+\frac{\sin(\pi \delta) \log_2 e}{\pi}\MeijerG*{2}{2}{2}{3}{0,~ 1-\delta}{0, 0, -\delta}{z}\quad\!\!\!\!\!\!,
\end{align}
where $s^*$ is the solution to the equation $s^{*^{\delta}}\Gamma(-\delta,s^*)=0$, given by lookup tables that are computed a priori for all relevant values of $\beta$. Furthermore, $\delta=\frac{2}{\beta}$, $D_\delta=\frac{s^*}{\log(1-\rm{sinc}(\delta))}$, $\operatorname{E_n}(x) = \int_{1}^{\infty} t^{-n}e^{-x t} dt $ is the exponential integral and
\begin{equation}
   \MeijerG*{m}{n}{p}{q}{a_1, \dots,a_n,a_{n+1},\dots, a_p}{b_1, \dots,b_m,b_{m+1},\dots, b_q}{z}  = \frac{1}{2\pi i} \int_{\mathcal{L}}^{} \frac{\prod_{k=1}^{m}\Gamma{(s+b_k)}\prod_{k=1}^{n}\Gamma{(1-a_k-s)}}{\prod_{k=n+1}^{p}\Gamma{(s+a_k)}\prod_{k=m+1}^{q}\Gamma{(1-b_k-s)}}z^{-s}ds\label{eqn:MeijerG},
\end{equation}
is the Meijer-G function. Improving the result of (\ref{eqn:Rate_Lozano}), the present paper introduces for the first time a closed form approximation for the ergodic rate in (\ref{eqn:Rate_Closed}), without the need for an a priori computed lookup table. Furthermore, the result of (\ref{eqn:Rate_Closed}) is significantly more tractable than the result of (\ref{eqn:Rate_Lozano}), and Fig. \ref{fig:RateLoz} demonstrates the superior performance of (\ref{eqn:Rate_Closed}) over (\ref{eqn:Rate_Lozano}) with respect to the tightness of the approximation. The accuracy and, more importantly, the tractability of the above expressions allow for the extension of the analysis to even more complex scenarios, as will be demonstrated in the following sections. 

\subsection{Ergodic Rate over Density of Users and BSs}
\label{ssec:RateDensity}

As already mentioned, the previous analysis corresponds to a scenario where the density of users is much greater than the density of BSs (i.e. $\lambda_{user}\gg \lambda$) and, therefore, every BS is in transmission mode. However, since in the envisaged UDNs the number of BSs is expected to be higher than the number of UE \cite{UDNs}, the analysis needs to be extended accordingly, taking into account the non transmitting mode of the excess BSs that do not have any UE in their coverage. In this course, the proposed tractable MGF is revised to account for the probability of the excess BSs to remain idle. This probability is defined by the density of UE $\lambda_{UE}$ and the density of BSs $\lambda$. Thus, following a similar approach as before, we derive closed form expressions for the DL ergodic capacity (i.e. peak and divided among intra-cell users) which depend on the density of UE $\lambda_{UE}$ and the density of BSs $\lambda$.  

The probability that a randomly chosen BS does not have any UE in its Voronoi cell and, therefore, goes into idle mode is given by \cite{Korean}:
\begin{align}
\label{eqn:P_inactive}
P_{inactive}&= \left(1+\frac{\lambda_{UE}}{3.5\lambda}\right)^{-3.5}
\end{align}
and the probability of a BS being in transmission mode and, thus, acting as an interferer in the DL, is denoted by:
\begin{align}
\label{eqn:P_active}
P_{active}&= 1-P_{inactive}.
\end{align}

Since only a subset of the BSs create interference in the DL, the density of the BSs $\lambda$ must be thinned out by the probability $P_{active}$. By introducing the thinned out density $\lambda P_{active}$ into (\ref{eqn:MGFinter}) an (\ref{eqn:MGF_Approx_Short}), the MGF is revised as follows:
\begin{align}
\label{eqn:MGFinter2}
	{M_{g}}\left( s; { L^{\left( 0 \right)} } \right)&= \exp \! \! \left(   \! \; {\int\limits_{{L^{\left( 0 \right)}} }^\infty {} \! \! \!  \left(\! {\exp \left( { - s\frac{{{P_{\rm{tx}}}}}{y}} \right)\!\!-\!\!1} \right)  \!  {\frac{{2\pi \lambda P_{active} }}{\beta }\left( \frac{1}{\kappa}\right)^{\frac{2}{\beta}}{\!\!\!\!y^{\frac{2}{\beta } - 1}}} dx}\!\! \right)\! \!
	\nonumber \\
	 & \mathop = \limits^{\left( a \right)} \exp  { \!\! {} \left( \! {\pi \lambda {{\left( {\frac{{ L^{\left( 0 \right)} }}{\kappa }} \right)}^{\frac{2}{\beta }}} \!\!\!\! \left( \! {1\!\!-\!\!\! {\;_1}{F_1} \! \!\left( { - \frac{2}{\beta },1 - \frac{2}{\beta },\frac{{ - s{P_{\rm{tx}}}}}{{ L^{\left( 0 \right)} }}} \! \right)} \! \! \right)} \! P_{active}\!\!\right) \!} \! \! \nonumber \\
& \mathop \approx \limits^{\left( b \right)} \begin{cases}
           \!  \exp   \left( \pi \lambda \left(\frac{L^{\left( 0 \right)}}{\kappa} \right)^\frac{2}{\beta}\!\! \! \left(\!\frac{-2 s P_{\rm{tx}}}{(\beta-2)L^{\left( 0 \right)}}\! + \!\frac{\left(s P_{\rm{tx}}\right)^2}{(2\beta-2)(L^{\left( 0 \right)})^2}\! \! \right)\! \!P_{active}\! \!\right)\!\!,& \!\!\!\frac{s P_{\rm{tx}}}{L^{\left( 0 \right)}}\!\! \leq \!\! c\\
           \! \exp \left(\pi \lambda  \left(\frac{L^{\left( 0 \right)}}{\kappa} \right)^\frac{2}{\beta}\! \!\left(\! \!-\! \!\left( \frac{s P_{\rm{tx}}}{L^{\left( 0 \right)}}\right)^{\frac{2}{\beta}} \! \!\Gamma\left(1-\frac{2}{\beta}\right)\! \!+\! \! 1 \right)\! \!P_{active}\! \! \right)\!\!,&\!\!\!\! \frac{s P_{\rm{tx}}}{L^{\left( 0 \right)}}\!\!>\!\! c
         \end{cases} \nonumber \\
& s \in \mathbb{R},
\end{align}
and, following the approach of (\ref{eqn:Pcov2}) for the MGF of (\ref{eqn:MGFinter2}), the coverage probability is now given by:
\begin{align}
\label{eqn:Pcov3}
\!\!\!\!\!\!\!\!\!\!P_{cov/active}& \mathop = \limits^{\left( a \right)} \frac{1}{1+ \left(\! {\;_1}{F_1}\left(-\frac{2}{\beta},1-\frac{2}{\beta},-\gamma\right) -1\right)P_{active}}\nonumber \\
\!\!\!\!\!\!\!\!\!\!& \mathop \approx \limits^{\left( b \right)}\frac{{\bf{1}} \left( \gamma \leq c \right)}{1+\left(-\frac{\gamma^2}{2\beta-2}+\frac{2\gamma}{\beta-2}\right)P_{active}}+\frac{{\bf{1}} \left( \gamma>c \right)}{1+\left(\gamma^{\frac{2}{\beta}}\Gamma(1-\frac{2}{\beta})-1\right)P_{active}}.
\end{align}

\begin{figure}[h!]
  \begin{center}
   \centerline{
    \includegraphics[height=9cm]{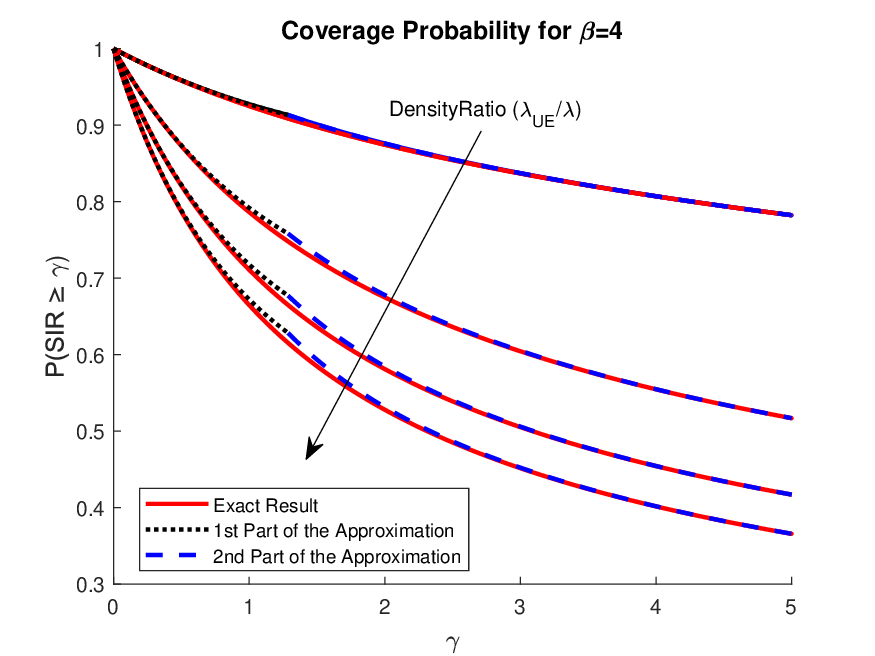}}
    \caption{Probability of coverage for different density ratios ($0.1\leq\lambda_{UE}/\lambda\leq1$). The density ratio increases in the direction of the arrow with a step of 0.3.}
    \label{fig:Pcov2}
  \end{center}
\end{figure}

The probability of coverage defined by the exact result of (\ref{eqn:Pcov3}a) and by the approximation of (\ref{eqn:Pcov3}b) is plotted in Fig. \ref{fig:Pcov2} for a path loss exponent $\beta=4$ and for density ratios $\lambda_{UE}/\lambda=0.1,0.4,0.7,1$, demonstrating, once again, the accuracy of the derived approximation.

Following the same approach as in (\ref{eqn:Rate}) for the results of (\ref{eqn:Pcov3}b), the DL ergodic peak rate can be computed in closed form as follows:
\begin{align}
\label{eqn:Rate2}
R_{peak}=& \int \limits_{0}^c \frac{1}{\left(1+\left(-\frac{w^2}{2\beta-2}+ \frac{2 w}{\beta-2} \right)P_{active}\right)(1+w)} dw \nonumber \\
&+\int \limits_{c}^\infty \frac{1}{\left(1+ \left(w^{\frac{2}{\beta}} \Gamma\left(1-\frac{2}{\beta} \right) -1\right)P_{active}\right)(1+w)} dw.
\end{align}
For path loss exponent values of $\beta=3, 4, 5$ the closed form expressions for the peak rate are given in Table \ref{tab:Expressions}. The expressions (\ref{eqn:Rate3Closed})-(\ref{eqn:Rate5Closed}) provide a closed form representation of the peak DL ergodic rate over the probability $P_{active}$ and, implicitly through (\ref{eqn:P_active}), over the density of BSs $\lambda$ and of UE $\lambda_{UE}$. 

\begin{figure}[h!]
  \begin{center}
   \centerline{
    \includegraphics[height=9cm]{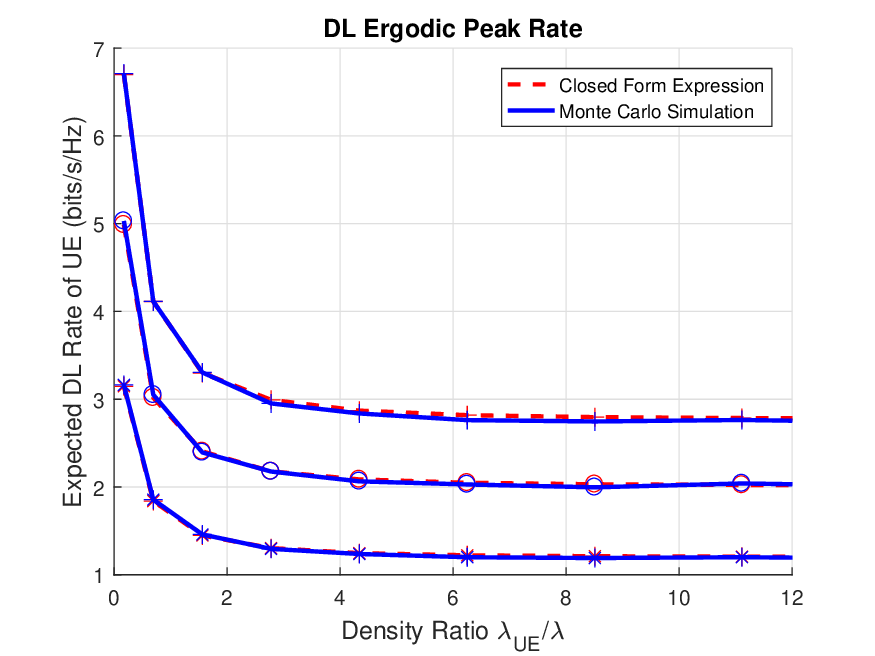}}
    \caption{DL ergodic peak rate vs ratio of densities $\lambda_{UE}/\lambda$ for a path loss exponent $\beta=$3$(*)$, 4$($o$)$ and 5$($+$)$.}
    \label{fig:RateUsersInstant}
  \end{center}
\end{figure}

\begin{figure}[h!]
  \begin{center}
   \centerline{
    \includegraphics[height=9cm]{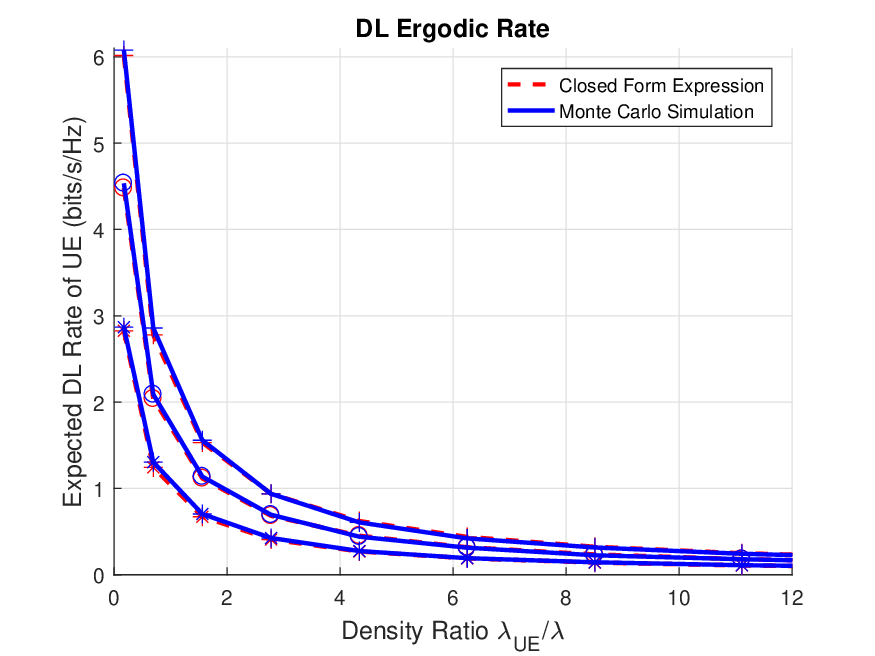}}
    \caption{DL ergodic capacity vs ratio of densities $\lambda_{UE}/\lambda$ for a path loss exponent  $\beta=$3$(*)$, 4$($o$)$ and 5$($+$)$.}
    \label{fig:RateUsers}
  \end{center}
\end{figure}

The probability that a randomly chosen UE is assigned a resource block at a given time and is served by its nearest BS is given by \cite{Korean}:
\begin{align}
\label{eqn:P_selection}
P_{selection}&= \frac{\lambda}{\lambda_{UE}}\left(1-(1+\frac{\lambda_{UE}}{3.5\lambda})^{-3.5}\right).
\end{align}
Employing this probability, a more sensible figure of merit than the peak rate can be derived. This figure of merit is the actual DL ergodic rate of the reference UE, i.e., the rate of the reference UE after dividing the available resources and, thus, the peak rate among all intra-cell UE. The latter is given by:
\begin{align}
\label{eqn:R_actual}
R&= R_{peak}P_{selection}.
\end{align}
The accuracy of expressions (\ref{eqn:R_actual})-(\ref{eqn:Rate5Closed}) is demonstrated in Fig. \ref{fig:RateUsersInstant} and \ref{fig:RateUsers} where the peak and the actual DL ergodic rates are plotted over the ratio of the densities $\lambda_{UE}/\lambda$ and compared against Monte Carlo simulations. 

In the Monte Carlo simulations, the BSs are deployed following the PPP $\Psi$ of density $\lambda$ and the users are deployed following the PPP $\Phi$ of density $\lambda_{UE}$, whereas the reference UE resides at the origin. BSs with no UE in their coverage (i.e. within their Voronoi cell) do not create any interference. To compute the actual rate, the number of users residing within the Voronoi cell of $BS_0$ are counted in each realization and the peak rate is divided among these users and the reference UE. The number of simulated BSs is fixed and the simulation area expands or contracts as the BS density $\lambda$ changes, in order to accommodate the predefined number of BSs, while the density of users $\lambda_{UE}$ remains fixed.  

The closed form expressions derived in (\ref{eqn:R_actual})-(\ref{eqn:Rate5Closed}) provide, among others, a substantial computational gain when compared to the computational time of the respective Monte Carlo simulations. Especially, since the simulation of a wireless network, that comprises both users and BSs of different spatial distributions, and the calculation of their relative distances, is computationally expensive. In order to demonstrate the gain arising by the employment of the closed form expressions of (\ref{eqn:R_actual})-(\ref{eqn:Rate5Closed}), the computational time of the derived closed form expressions is tabulated in Table \ref{tab:Computation}. Since this time is not an absolute metric, but depends on the hardware employed, Table \ref{tab:Computation} presents the computational time of the expressions as a percentage of the computational time required for simulating the respective wireless networks of Fig. \ref{fig:RateUsers} using the same hardware.
\begin{table}[h]
\caption{Computational time of expressions (\ref{eqn:R_actual})-(\ref{eqn:Rate5Closed}) as a percentage of the respective computational time for simulating the scenarios of Fig. \ref{fig:RateUsers}.}
\label{tab:Computation}
\begin{center}
\begin{tabular}{|c|c|c|c|c|}
  \hline
  \multirow{2}{*}{\bf Path-loss ($\beta$)}& \multicolumn{4}{c|}{\bf Density Ratio $\lambda_{UE}/\lambda$}\\ \cline{2-5}
 &0.17&4.34&8.51&11.11\\ \hline  \hline
   3&$6.9~10^{-4}~\%$&$6.7~10^{-4}~\%$&$3.4~10^{-7}~\%$&$2.7~10^{-7}~\%$\\ \hline
   4&$6.9~10^{-4}~\%$&$6.8~10^{-4}~\%$&$1.8~10^{-7}~\%$&$1.3~10^{-7}~\%$\\ \hline
   5&$2.2~10^{-3}~\%$&$4.6~10^{-3}~\%$&$2.3~10^{-6}~\%$&$1.6~10^{-6}~\%$\\
 \hline
 \end{tabular}
\end{center}
\end{table}

Table \ref{tab:Computation} demonstrates that the computational gain arising by the derived expressions is immense. The time required for the analytical computation of the rate is practically zero compared to the time required for simulating the respective scenario. At this point it should also be noted, that the variations in the values of the computational time arise due to the variations in the time required for the respective simulations. That is since the simulation area has to expand and contract as $\lambda$ changes in order to accommodate a fixed number of BSs. The variations with respect to $\beta$ emerge due to the different complexity of the analytical expressions of (\ref{eqn:Rate3Closed})-(\ref{eqn:Rate5Closed}), with respect to $\beta$.

Apart from the computational gain that has been demonstrated by Table \ref{tab:Computation}, the closed form expressions of (\ref{eqn:R_actual})-(\ref{eqn:Rate5Closed}) and the respective figures, i.e. Fig. \ref{fig:RateUsersInstant} and \ref{fig:RateUsers} provide a deep understanding of the behavior of UDNs as the density of users $\lambda_{UE}$ and BSs $\lambda$ changes. First of all, expressions (\ref{eqn:Rate3Closed})-(\ref{eqn:Rate5Closed}) and Fig. \ref{fig:RateUsersInstant} verify one of the fundamental findings of stochastic geometry, already explained in Section \ref{sec:introduction}, that in the fully loaded case (i.e.  for $\lambda_{UE}\gg \lambda$) the ergodic rate remains invariant while the BS density changes. However, Fig. \ref{fig:RateUsers} demonstrates that even if the peak rate remains invariant, the rate of the users that have to share this peak rate tends to zero as the density ratio (i.e. the expected number of users per typical cell) increases. This behavior demonstrates why the envisaged UDNs are expected to comprise more BSs than users, highlighting the importance of the non-fully loaded case.  Given the importance of the non-fully loaded case, expressions (\ref{eqn:Rate3Closed})-(\ref{eqn:Rate5Closed}) and Fig. \ref{fig:RateUsersInstant} allow for the first time to quantify the threshold between the non-fully and the fully loaded case. In particular, in all three expressions for the different values of $\beta$, the network exhibits the behavior of a fully loaded network for $\lambda_{UE}/ \lambda>4$, way earlier than implied by the notation $\lambda_{UE}\gg \lambda$. 

Another interesting finding can be derived from the behavior of the network in the non-fully loaded case, when the BSs that do not comprise any user in their coverage do not create any interference. In this setup, Fig. \ref{fig:RateUsersInstant} demonstrates that the densification of the network can provide substantial capacity gains, since the achieved DL ergodic rate in this range is significantly higher than the rate achieved at the fully loaded case. Thus,  Fig. \ref{fig:RateUsersInstant} demonstrates that just by switching off the BSs that do not comprise users in their coverage, substantial capacity gains can be engendered for the network. This motivates the use of even more intricate schemes where BSs can be switched off strategically to mitigate interference. This optimization problem is only one of the problems that the derived expressions can be applied to, with additional applications being presented in the following section.

Moreover, expressions (\ref{eqn:Rate3Closed})-(\ref{eqn:Rate5Closed}) and the respective figures, verify the results of the expressions (\ref{eqn:Rate_Closed}) and (\ref{eqn:Rate_Lozano}), and of the figures \ref{fig:Rate} and \ref{fig:RateLoz}, according to which the DL ergodic rate increases monotonically with the path-loss exponent $\beta$. In particular, since the distance from the UE to the serving BS is always smaller than the distance to the interfering BSs, i.e. since $\left\lVert x_i \right\rVert>\left\lVert x_0 \right\rVert$, each term of the sum $\sum\limits_{\substack{x \in \Psi, \\ x \neq x_0}}\left(\frac{\left\lVert x_0 \right\rVert}{\left\lVert x_i \right\rVert}\right)^\beta$ decreases with $\beta$. Consequently, the SIR and the DL ergodic rate increase monotonically with $\beta$.  

\subsection{Applications of the Derived Expressions}
\label{ssec:Applications}

Figures \ref{fig:RateUsersInstant} and \ref{fig:RateUsers} corroborate the accuracy of the derived expressions, while providing a deeper understanding of the network performance as a function of the user and BS densities. Hence, these expressions, that for the first time associate the DL ergodic rate with the densities of the UE and of the BSs is a closed form, pave the way for the investigation of complex optimization problems, toward improving UDN operation and offered QoS. 

In particular, given the maximum density of UE $\lambda_{UE}$ during the operation of the network and the minimum rate requirement per user (imposed by the QoS constraints), Fig. \ref{fig:RateUsers} and (\ref{eqn:R_actual}) can be employed by network operators to define the minimum BS density $\lambda$ that guarantees this rate. Hence, (\ref{eqn:R_actual}) can be employed as a rule of thumb for the lower limit of the densification of BSs that guarantees the QoS objectives, implicitly quantifying the minimum capital expenditure required by network operators. 

In addition to that, given the aforementioned minimum rate requirement per user and a network of BS density $\lambda$, comprising UE whose density $\lambda_{UE}$ varies during the operation of the network, Fig. \ref{fig:RateUsers} and (\ref{eqn:R_actual}) can be employed to dynamically define the probability of transmission $P_{active}$ that achieves the predefined rate requirement as $\lambda_{UE}$ changes. This probability can indicate the density of transmitting BSs $\lambda P_{active}$, i.e. the density of the BSs comprising at least one UE in their coverage. The latter density can be input into optimization modules, to be used as a starting point for the search aiming at pinpointing the optimum set of BSs to be switched off, additionally to those that do not comprise any UE in their coverage. Given the high density of the BSs, strategically switching off the best serving BSs of a UE for the latter to be served by a neighboring BS has a only minimal impact on the path-loss, while it can provide substantial capacity gains through the mitigation of the interference. 

\section{Conclusions}
\label{sec:Conclusions}
The present paper has demonstrated how stochastic geometry tools can be exploited to derive not just exact but cumbersome expressions, but also simple, albeit extremely accurate closed form expressions that allow for the investigation of complex optimization problems. The resolution of these problems is essential in order to reap the capacity benefits of UDNs. In this direction, the present paper presented an accurate and tractable approximation for the MGF of the aggregate other-cell interference. Given the pivotal role of the MGF in stochastic geometry analyses, the derived approximation can be employed by a multitude of applications to simplify the analysis and facilitate the derivation of closed form expressions. 

Building upon this result, the present paper has focused on the interference limited case, providing very tractable expressions for the coverage probability, as well as closed form expressions for the DL ergodic capacity (i.e. peak and actual) which depend on the density of UE $\lambda_{UE}$ and the density of BSs $\lambda$. The derived expressions and their dependence on the user and BS densities set out a densification road map for network operators and designers of significant practical and commercial value.  Moreover, such expressions can be used for the resolution of complex optimization problems in real time, improving the network performance 
and operation.

\section*{\begin{scriptsize}List of Abbreviations\end{scriptsize}}
\begin{scriptsize}
BS: base station; CCDF: complementary cumulative distribution function; DL: downlink; MC: macro cell; MGF: moment generating function; LOS: line of sight; PDF: probability density function; PGFL: probability generating fuctional; PPP: Poisson point process; QoS: quality of service; SINR: signal to interference plus noise ratio; SIR: signal to interference ratio; UDN: ultra dense network; UE: user equipment; UL: uplink.
\end{scriptsize}

\section*{\begin{scriptsize}Availability of data and materials\end{scriptsize}}
\begin{scriptsize}
Not applicable. No datasets were imported or generated while developing the presented material. 
\end{scriptsize}

\section*{\begin{scriptsize}Competing interests\end{scriptsize}}
\begin{scriptsize}
The authors declare that they have no competing interests.
\end{scriptsize}

\section*{\begin{scriptsize}Funding\end{scriptsize}}
\begin{scriptsize}
The work presented in the present paper has been carried out within the framework of the project ETN-5Gwireless (this project has received funding from the European Union’s Horizon 2020 research and innovation programme under the Marie Sk\l{}odowska-Curie grant agreement No. 641985). Moreover, the work has been partially funded through the grant 2017 SGR 578 (funded by the Catalan Government—Secretaria d'Universitats i Recerca, Departament d'Empresa i Coneixement, Generalitat de Catalunya, AGAUR) and the project TEC2016-77148-C2-1-R (AEI/FEDER, UE): 5G\&B-RUNNER-UPC (funded by the Agencia Estatal de Investigacion, AEI, and Fondo Europeo de Desarrollo Regional, FEDER).
\end{scriptsize}

\section*{\begin{scriptsize}Authors’ contributions\end{scriptsize}}
\begin{scriptsize}
All authors have contributed to this research work. Moreover, all authors have read and approved the final manuscript.
\end{scriptsize}

\balance 




\begin{thebibliography}{20}
\ifx \bisbn   \undefined \def \bisbn  #1{ISBN #1}\fi
\ifx \binits  \undefined \def \binits#1{#1}\fi
\ifx \bauthor  \undefined \def \bauthor#1{#1}\fi
\ifx \batitle  \undefined \def \batitle#1{#1}\fi
\ifx \bjtitle  \undefined \def \bjtitle#1{#1}\fi
\ifx \bvolume  \undefined \def \bvolume#1{\textbf{#1}}\fi
\ifx \byear  \undefined \def \byear#1{#1}\fi
\ifx \bissue  \undefined \def \bissue#1{#1}\fi
\ifx \bfpage  \undefined \def \bfpage#1{#1}\fi
\ifx \blpage  \undefined \def \blpage #1{#1}\fi
\ifx \burl  \undefined \def \burl#1{\textsf{#1}}\fi
\ifx \doiurl  \undefined \def \doiurl#1{\textsf{#1}}\fi
\ifx \betal  \undefined \def \betal{\textit{et al.}}\fi
\ifx \binstitute  \undefined \def \binstitute#1{#1}\fi
\ifx \binstitutionaled  \undefined \def \binstitutionaled#1{#1}\fi
\ifx \bctitle  \undefined \def \bctitle#1{#1}\fi
\ifx \beditor  \undefined \def \beditor#1{#1}\fi
\ifx \bpublisher  \undefined \def \bpublisher#1{#1}\fi
\ifx \bbtitle  \undefined \def \bbtitle#1{#1}\fi
\ifx \bedition  \undefined \def \bedition#1{#1}\fi
\ifx \bseriesno  \undefined \def \bseriesno#1{#1}\fi
\ifx \blocation  \undefined \def \blocation#1{#1}\fi
\ifx \bsertitle  \undefined \def \bsertitle#1{#1}\fi
\ifx \bsnm \undefined \def \bsnm#1{#1}\fi
\ifx \bsuffix \undefined \def \bsuffix#1{#1}\fi
\ifx \bparticle \undefined \def \bparticle#1{#1}\fi
\ifx \barticle \undefined \def \barticle#1{#1}\fi
\ifx \bconfdate \undefined \def \bconfdate #1{#1}\fi
\ifx \botherref \undefined \def \botherref #1{#1}\fi
\ifx \url \undefined \def \url#1{\textsf{#1}}\fi
\ifx \bchapter \undefined \def \bchapter#1{#1}\fi
\ifx \bbook \undefined \def \bbook#1{#1}\fi
\ifx \bcomment \undefined \def \bcomment#1{#1}\fi
\ifx \oauthor \undefined \def \oauthor#1{#1}\fi
\ifx \citeauthoryear \undefined \def \citeauthoryear#1{#1}\fi
\ifx \endbibitem  \undefined \def \endbibitem {}\fi
\ifx \bconflocation  \undefined \def \bconflocation#1{#1}\fi
\ifx \arxivurl  \undefined \def \arxivurl#1{\textsf{#1}}\fi
\csname PreBibitemsHook\endcsname

\bibitem{History}
\begin{barticle}
\bauthor{\bsnm{Chandrasekhar}, \binits{V.}},
\bauthor{\bsnm{Andrews}, \binits{J.G.}},
\bauthor{\bsnm{Gatherer}, \binits{A.}}:
\batitle{Femtocell networks: a survey}.
\bjtitle{IEEE Communications Magazine}
\bvolume{46}(\bissue{9}),
\bfpage{59}--\blpage{67}
(\byear{2008})
\end{barticle}
\endbibitem

\bibitem{Baccelli}
\begin{barticle}
\bauthor{\bsnm{Baccelli}, \binits{F.}},
\bauthor{\bsnm{Blaszczyszyn}, \binits{B.}},
\bauthor{\bsnm{Muhlethaler}, \binits{P.}}:
\batitle{An {A}loha protocol for multihop mobile wireless networks}.
\bjtitle{IEEE Transactions on Information Theory}
\bvolume{52},
\bfpage{421}--\blpage{436}
(\byear{2006})
\end{barticle}
\endbibitem

\bibitem{Haenggi}
\begin{bbook}
\bauthor{\bsnm{Haenggi}, \binits{M.}}:
\bbtitle{Stochastic Geometry for Wireless Networks}.
\bpublisher{Cambridge University Press},
\blocation{Cambridge}
(\byear{2013})
\end{bbook}
\endbibitem

\bibitem{Andrews1}
\begin{barticle}
\bauthor{\bsnm{Andrews}, \binits{J.G.}},
\bauthor{\bsnm{Baccelli}, \binits{F.}},
\bauthor{\bsnm{Ganti}, \binits{R.K.}}:
\batitle{A tractable approach to coverage and rate in cellular networks}.
\bjtitle{IEEE Transactions on Communications}
\bvolume{59}(\bissue{11}),
\bfpage{3122}--\blpage{3134}
(\byear{2011})
\end{barticle}
\endbibitem

\bibitem{Andrews2}
\begin{barticle}
\bauthor{\bsnm{Dhillon}, \binits{H.S.}},
\bauthor{\bsnm{Ganti}, \binits{R.K.}},
\bauthor{\bsnm{Baccelli}, \binits{F.}},
\bauthor{\bsnm{Andrews}, \binits{J.G.}}:
\batitle{Modeling and analysis of k-tier downlink heterogeneous cellular
  networks}.
\bjtitle{IEEE Journal on Selected Areas in Communications}
\bvolume{30}(\bissue{3}),
\bfpage{550}--\blpage{560}
(\byear{2012})
\end{barticle}
\endbibitem

\bibitem{TWC}
\begin{barticle}
\bauthor{\bsnm{Singh}, \binits{S.}},
\bauthor{\bsnm{Zhang}, \binits{X.}},
\bauthor{\bsnm{Andrews}, \binits{J.G.}}:
\batitle{Joint rate and {SINR} coverage analysis for decoupled uplink-downlink
  biased cell associations in {H}et{N}ets}.
\bjtitle{IEEE Transactions on Wireless Communications}
\bvolume{14}(\bissue{10}),
\bfpage{5360}--\blpage{5373}
(\byear{2015})
\end{barticle}
\endbibitem

\bibitem{Marco1}
\begin{barticle}
\bauthor{\bsnm{Renzo}, \binits{M.D.}},
\bauthor{\bsnm{Lu}, \binits{W.}},
\bauthor{\bsnm{Guan}, \binits{P.}}:
\batitle{The intensity matching approach: A tractable stochastic geometry
  approximation to system-level analysis of cellular networks}.
\bjtitle{IEEE Transactions on Wireless Communications}
\bvolume{15}(\bissue{9}),
\bfpage{5963}--\blpage{5983}
(\byear{2016})
\end{barticle}
\endbibitem

\bibitem{Lopez1}
\begin{barticle}
\bauthor{\bsnm{Ding}, \binits{M.}},
\bauthor{\bsnm{Wang}, \binits{P.}},
\bauthor{\bsnm{Lopez-Perez}, \binits{D.}},
\bauthor{\bsnm{Mao}, \binits{G.}},
\bauthor{\bsnm{Lin}, \binits{Z.}}:
\batitle{Performance impact of los and nlos transmissions in dense cellular
  networks}.
\bjtitle{IEEE Transactions on Wireless Communications}
\bvolume{15}(\bissue{3}),
\bfpage{2365}--\blpage{2380}
(\byear{2016})
\end{barticle}
\endbibitem

\bibitem{UDNs}
\begin{barticle}
\bauthor{\bsnm{Andrews}, \binits{J.G.}}:
\batitle{Seven ways that hetnets are a cellular paradigm shift}.
\bjtitle{IEEE Communications Magazine}
\bvolume{51}(\bissue{3}),
\bfpage{136}--\blpage{144}
(\byear{2013})
\end{barticle}
\endbibitem

\bibitem{Lopez2}
\begin{barticle}
\bauthor{\bsnm{Ding}, \binits{M.}},
\bauthor{\bsnm{Lopez-Perez}, \binits{D.}},
\bauthor{\bsnm{Mao}, \binits{G.}},
\bauthor{\bsnm{Lin}, \binits{Z.}}:
\batitle{Performance impact of idle mode capability on dense small cell
  networks}.
\bjtitle{IEEE Transactions on Vehicular Technology}
\bvolume{66}(\bissue{11}),
\bfpage{10446}--\blpage{10460}
(\byear{2017})
\end{barticle}
\endbibitem

\bibitem{VTC}
\begin{bchapter}
\bauthor{\bsnm{Aravanis}, \binits{A.I.}},
\bauthor{\bsnm{Munoz}, \binits{O.}},
\bauthor{\bsnm{Pascual-Iserte}, \binits{A.}},
\bauthor{\bsnm{Di-Renzo}, \binits{M.}}:
\bctitle{On the coordination of base stations in ultra dense cellular
  networks}.
In: \bbtitle{2019 IEEE 88th Vehicular Technology Conference (VTC Spring)}
(\byear{accepted for publication})
\end{bchapter}
\endbibitem

\bibitem{Stoyan}
\begin{bbook}
\bauthor{\bsnm{Chiu}, \binits{S.N.}},
\bauthor{\bsnm{Stoyan}, \binits{D.}},
\bauthor{\bsnm{Kendall}, \binits{W.S.}},
\bauthor{\bsnm{Mecke}, \binits{J.}}:
\bbtitle{Stochastic Geometry and Its Applications}.
\bpublisher{John Wiley and Sons},
\blocation{West Sussex}
(\byear{2013})
\end{bbook}
\endbibitem

\bibitem{Geordie}
\begin{botherref}
\oauthor{\bsnm{Geordie}, \binits{G.}}:
Device-to-device communication and wearable networks: Harnessing spatial
  proximity.
PhD thesis,
Universitat Pompeu Fabra
(2017)
\end{botherref}
\endbibitem

\bibitem{Thanh}
\begin{botherref}
\oauthor{\bsnm{Lam-Thanh}, \binits{T.U.}}:
New methods for the analysis and optimization of cellular networks by using
  stochastic geometry.
PhD thesis,
Universite Paris-Saclay
(2018)
\end{botherref}
\endbibitem

\bibitem{HyperGeometric}
\begin{barticle}
\bauthor{\bsnm{Renzo}, \binits{M.D.}},
\bauthor{\bsnm{Lu.}, \binits{W.}}:
\batitle{System-level analysis/optimization of cellular networks with
  simultaneous wireless information and power transfer: Stochastic geometry
  modeling}.
\bjtitle{IEEE Trans. Vehicular Technology}
\bvolume{66},
\bfpage{2251}--\blpage{2275}
(\byear{2017})
\end{barticle}
\endbibitem

\bibitem{MarkedPPP}
\begin{bchapter}
\bauthor{\bsnm{Lee}, \binits{J.}},
\bauthor{\bsnm{Baccelli}, \binits{F.}}:
\bctitle{On the effect of shadowing correlation on wireless network
  performance}.
In: \bbtitle{IEEE INFOCOM 2018 - IEEE Conference on Computer Communications},
pp. \bfpage{1601}--\blpage{1609}
(\byear{2018})
\end{bchapter}
\endbibitem

\bibitem{PDF}
\begin{barticle}
\bauthor{\bsnm{Sadr}, \binits{S.}},
\bauthor{\bsnm{Adve}, \binits{R.S.}}:
\batitle{Partially-distributed resource allocation in small-cell networks}.
\bjtitle{IEEE Transactions on Wireless Communications}
\bvolume{13}(\bissue{12}),
\bfpage{6851}--\blpage{6862}
(\byear{2014})
\end{barticle}
\endbibitem

\bibitem{Marco2}
\begin{barticle}
\bauthor{\bsnm{Renzo}, \binits{M.D.}},
\bauthor{\bsnm{Guidotti}, \binits{A.}},
\bauthor{\bsnm{Corazza}, \binits{G.E.}}:
\batitle{Average rate of downlink heterogeneous cellular networks over
  generalized fading channels: A stochastic geometry approach.}
\bjtitle{IEEE Transactions on Communications}
\bvolume{61}(\bissue{7}),
\bfpage{3050}--\blpage{3071}
(\byear{2013})
\end{barticle}
\endbibitem

\bibitem{Lozano}
\begin{botherref}
\oauthor{\bsnm{Geordie}, \binits{G.}},
\oauthor{\bsnm{Mungara}, \binits{R.K.}},
\oauthor{\bsnm{Lozano}, \binits{A.}},
\oauthor{\bsnm{Haenggi}, \binits{M.}}:
Ergodic spectral efficiency in {MIMO} cellular networks.
IEEE Transactions on Wireless Communications
\textbf{16}(5)
(2017)
\end{botherref}
\endbibitem

\bibitem{Korean}
\begin{bchapter}
\bauthor{\bsnm{Yu}, \binits{S.M.}},
\bauthor{\bsnm{Kim}, \binits{S.L.}}:
\bctitle{Downlink capacity and base station density in cellular networks}.
In: \bbtitle{2013 11th International Symposium and Workshops on Modeling and
  Optimization in Mobile, Ad Hoc and Wireless Networks (WiOpt)},
pp. \bfpage{119}--\blpage{124}
(\byear{2013})
\end{bchapter}
\endbibitem

\end{thebibliography}

\begin{table*}[t]
\caption{Closed Form Expressions for the DL Ergodic Peak Rate for different path loss exponents}
\label{tab:Expressions}
 \begin{adjustwidth}{-2cm}{}

\begin{tabular}{|c|c|}
  \hline
  {\bf $\beta$} & {\bf DL Ergodic Peak Rate:} \\ \hline
 \hline
  3 &  \vbox{\begin{align}
\label{eqn:Rate3Closed}
&R_{peak}=\nonumber \\ 
& \frac{-4}{P_{active}}\left(\frac{\log\left(\frac{c-4+b}{-4+b}\right)}{2b(-5+b)}+\frac{\log\left(\frac{c-4-b}{-4-b}\right)}{2b(5+b)}-\frac{\log\left(c+1\right)}{(-5+b)(5+b)}\right) -\frac{\sqrt{3}P_{active}\arctan\left({\frac{-1+2c^{1/3}}{\sqrt{3}}}\right)\Gamma({\frac{1}{3}})}{1+P_{active}(-2+P_{active}+(P_{active}-1)\Gamma({\frac{1}{3}})+P_{active}\left(\Gamma({\frac{1}{3}})\right)^2)}\nonumber \\
&+\Bigg[-(1-P_{active})^{3/2}P_{active}~\pi\left(-\sqrt{3}+3\sqrt{-\frac{P_{active}\Gamma(\frac{1}{3})}{P_{active}-1}}\right)\Gamma\left(\frac{1}{3}\right)-6\left(P_{active}-1\right)\left(P_{active}\Gamma\left(\frac{1}{3}\right)\right)^{3/2}\nonumber \\
& \arctan{\left(\sqrt{\frac{P_{active}\Gamma\left(\frac{1}{3}\right)}{1-P_{active}}}c^{1/3}\right)}+\sqrt{1-P_{active}}\Bigg[\sqrt3 \left(P_{active}\Gamma\left(\frac{1}{3}\right)\right)^2 \pi -2P_{active}\Gamma\left(\frac{1}{3}\right)\left(-1+P_{active}\left(1+\Gamma\left(\frac{1}{3}\right)\right)\right)\nonumber \\
&\log\left(1+c^{1/3}\right)+P_{active}\Gamma\left(\frac{1}{3}\right)\left(-1+P_{active}\left(1+\Gamma\left(\frac{1}{3}\right)\right)\right)\log\left(1-c^{1/3}+c^{2/3}\right)+\left(-1+P_{active}\right)^2\bigg[-2\log\left(1+c\right)\nonumber \\
&-3\log\left(P_{active}\Gamma\left(\frac{1}{3}\right)\right)+3\log\left(1-P_{active}\left(-1+c^{2/3}\Gamma\left(\frac{1}{3}\right)\right)\right)\bigg]
\Bigg]\Bigg]/\nonumber \\
&\Bigg[2\sqrt{1-P_{active}}\left(-(-1+P_{active})^3+\left(P_{active}\Gamma\left(\frac{1}{3}\right)\right)^3\right)\Bigg],
where~b=2\sqrt{4+\frac{1}{P_{active}}},~ c=1.2528
\end{align}} \\  \hline
  4 &  \vbox{\begin{align}
\label{eqn:Rate4Closed}
&R_{peak}= \nonumber \\
&\frac{-6}{P_{active}}\left(\frac{\log\left(\frac{c-3+b}{-3+b}\right)}{2b(-4+b)}+\frac{\log\left(\frac{c-3-b}{-3-b}\right)}{2b(4+b)}-\frac{\log\left(c+1\right)}{(-4+b)(4+b)}\right)+\frac{-2\log(P_{active})(1+P_{active})+(P_{active}-1)\log\left((1+c)\pi\right)}{1+P_{active}(-2+P_{active}(1+\pi))} \nonumber \\
&+\frac{\pi^{1.5}P_{active}-2\sqrt{\pi}P_{active}\arctan{\sqrt{c}}-2(P_{active}-1)\log\left(1-P_{active}+\sqrt{\pi c}P_{active}\right)}{1+P_{active}(-2+P_{active}(1+\pi))},where~b=\sqrt{9+\frac{6}{P_{active}}},~ c=1.2873
\end{align}} \\  \hline
  5 &  \vbox{\begin{align}
&R_{peak}= \nonumber \\
& \frac{-8}{P_{active}}\left(\frac{3\log\left(\frac{c-8/3+b}{-8/3+b}\right)}{2b(-11+3b)}+\frac{3\log\left(\frac{c-8/3-b}{-8/3-b}\right)}{2b(11+3b)}-\frac{9\log\left(c+1\right)}{(-11+3b)(11+3b)}\right)+\Bigg[10\sqrt{\frac{1}{1-P_{active}}}\left(-1+P_{active}\right)^2\nonumber \\
& \pi \left(P_{active}\Gamma\left(\frac{3}{5}\right)\right)^{5/2}-20\left(1-P_{active}\right)^{3/2} \left(P_{active}\Gamma\left(\frac{3}{5}\right)\right)^{5/2}\arctan{\left(\frac{\left(P_{active}\Gamma\left(\frac{3}{5}\right)\right)^{1/2}c^{1/5}}{\sqrt{1-P_{active}}}\right)}+2\sqrt{2}\left(P_{active}\Gamma\left(\frac{3}{5}\right)\right) \nonumber \\
&\arctan\left(\frac{1+\sqrt{5}-4c^{1/5}}{\sqrt{10-2\sqrt{5}}}\right)\left(1+P_{active}\left(-1+\Gamma\left(\frac{3}{5}\right) \right)\right)\Bigg(\sqrt{5+\sqrt{5}}+P_{active}\Bigg(\left(-2+P_{active}\left(1+\left(\Gamma\left(\frac{3}{5}\right)\right)^2\right)\right)\nonumber \\
&\sqrt{5+\sqrt{5}}+\sqrt{10+4\sqrt{5}}\left(-1+P_{active}\right)\Gamma\left(\frac{3}{5}\right)\Bigg)\Bigg)+\left(-1+P_{active}\right)\left(P_{active}\Gamma\left(\frac{3}{5}\right)\right)^{3}\left(\sqrt{10-2\sqrt{5}}\pi-\log\left(4\right)\right)\nonumber \\
&+\left(P_{active}\Gamma\left(\frac{3}{5}\right)\right)^4\left(\sqrt{2\left(5+\sqrt{5}\right)}\pi-\log(4)\right)\nonumber
\end{align}}\\ \hline 
\end{tabular}
\end{adjustwidth}
\end{table*}

\begin{table*}[t]
 \begin{adjustwidth}{-2cm}{}
\begin{tabular}{|c|c|}
\hline
  5 &  \vbox{\begin{align}
\label{eqn:Rate5Closed}
&-\left(P_{active}\Gamma\left(\frac{3}{5}\right)\right)^2 \Bigg(\sqrt{10-2\sqrt{5}}\left(-1+P_{active}\right)^2\pi+\log(4)\left(1+\left(P_{active}\right)^2\right)-P_{active}\log(16)\Bigg)\nonumber \\
&-\left(P_{active}-1\right)\left(P_{active}\Gamma\left(\frac{3}{5}\right)\right)\left(\sqrt{2(5+\sqrt{5})}\left(P_{active}-1\right)^2\pi+\log(4)\left(1+\left(P_{active}\right)^2\right)-P_{active}\log(16)\right)\nonumber \\
&-4\log(1+c)-10\log\left(P_{active}\Gamma\left(\frac{3}{5}\right)\right)+P_{active}\bigg[\rm{i}\left(1+P_{active}\left(-1+\Gamma\left(\frac{3}{5}\right)\right)\right)\Gamma\left(\frac{3}{5}\right)\nonumber \\
&\bigg(\sqrt{10-2\sqrt{5}}+P_{active}\bigg(-2\sqrt{10-2\sqrt{5}}+\sqrt{10-2\sqrt{5}}P_{active}-2\sqrt{5-2\sqrt{5}}\left(-1+P_{active}\right)\Gamma\left(\frac{3}{5}\right)\nonumber \\
&+\sqrt{10-2\sqrt{5}}P_{active}\Gamma\left(\frac{3}{5}\right)^2 \bigg)\bigg)\log\left(-1+\frac{\rm{i}\left(-1+\sqrt{5}+4c^{1/5}\right)}{\sqrt{2\left(5+\sqrt{5}\right)}}\right)+P_{active}\Gamma\left(\frac{3}{5}\right)^4\Bigg(-\rm{i}\sqrt{10-2\sqrt{5}}\nonumber \\
&\log\left(1+\frac{\rm{i}(-1+\sqrt{5}+4c^{1/5})}{\sqrt{2(5+\sqrt{5})}}\right)-4\log\left(1+c^{1/5}\right)+(1+\sqrt{5})\log\left(2+(-1+\sqrt{5})c^{1/5}+2c^{2/5}\right)\nonumber \\
&-(-1+\sqrt{5})\log\left(2-(1+\sqrt{5})c^{1/5}+2c^{2/5}\right)\Bigg)+\left(-1+P_{active}\right)^3\Gamma\left(\frac{3}{5}\right)\Bigg(\rm{i}\sqrt{10-2\sqrt{5}}\nonumber  \\
&\log\left(1+\frac{\rm{i}(-1+\sqrt{5}+4c^{1/5})}{\sqrt{2(5+\sqrt{5})}}\right)-4\log\left(1+c^{1/5}\right)+(1+\sqrt{5})\log\left(2+(-1+\sqrt{5})c^{1/5}+2c^{2/5}\right)\nonumber \\
&-(-1+\sqrt{5})\log\left(2-(1+\sqrt{5})c^{1/5}+2c^{2/5}\right)\Bigg)+\left(\left(1+P_{active}\right)\Gamma\left(\frac{3}{5}\right)\right)^2P_{active}\Bigg(-\rm{i}\sqrt{2(5+\sqrt{5})}\nonumber \\
&\log\left(1+\frac{\rm{i}(-1+\sqrt{5}+4c^{1/5})}{\sqrt{2(5+\sqrt{5})}}\right)-4\log\left(1+c^{1/5}\right)-(-1+\sqrt{5})\log\left(2+(-1+\sqrt{5})c^{1/5}+2c^{2/5}\right)\nonumber \\
&(1+\sqrt{5})\log\left(2-(1+\sqrt{5})c^{1/5}+2c^{2/5}\right)\Bigg)+\left(-1+P_{active}\right)\left(P_{active}\right)^2\Gamma\left(\frac{3}{5}\right)^3\Bigg(\rm{i}\sqrt{2(5+\sqrt{5})}\nonumber \\
&\log\left(1+\frac{\rm{i}(-1+\sqrt{5}+4c^{1/5})}{\sqrt{2(5+\sqrt{5})}}\right)-4\log\left(1+c^{1/5}\right)+(-1+\sqrt{5})\log\left(2+(-1+\sqrt{5})c^{1/5}+2c^{2/5}\right)\nonumber \\
&+(1+\sqrt{5})\log\left(2-(1+\sqrt{5})c^{1/5}+2c^{2/5}\right)\Bigg)-2\left(-2+P_{active}\right)\left(2+P_{active}\left(-2+P_{active}\right)\right)\nonumber \\
&\left(2\log\left(1+c\right)+5\log\left(P_{active}\Gamma\left(\frac{3}{5}\right)\right)-5\log\left(1-P_{active}+P_{active}c^{2/5}\Gamma\left(\frac{3}{5}\right)\right)\right)\bigg]\nonumber \\
&+10\log\left(1-P_{active}+P_{active}c^{2/5}\Gamma\left(\frac{3}{5}\right)\right) \Bigg]/\left(-4\left(-1+P_{active}\right)^5+4\left(P_{active}\right)^5\Gamma\left(\frac{3}{5}\right)^5\right),\nonumber \\
&where~b=2\sqrt{\frac{16}{9}+\frac{2}{P_{active}}},~ c=1.3099
\end{align}}\\ \hline 
\end{tabular}
\end{adjustwidth}
\end{table*}




%




\end{document}